\documentclass[preprint,psfig]{aastex}

\usepackage[onecolumn]{emulateapj5}

\def\ltsima{$\; \buildrel < \over \sim \;$}

\def\lsim{\lower.5ex\hbox{\ltsima}}
\def\gtsima{$\; \buildrel > \over \sim \;$}
\def\gsim{\lower.5ex\hbox{\gtsima}}
\begin{document}
\title{The Extended Starformation History of the First Generation of Stars, and the Reionization of Cosmic Hydrogen}

\author{J. Stuart B. Wyithe\altaffilmark{1} and Renyue Cen\altaffilmark{2}}

\email{swyithe@physics.unimelb.edu.au, cen@astro.princeton.edu}

\altaffiltext{1}{The University of Melbourne, Parkville, 3010, Australia}
\altaffiltext{2}{Princeton University Observatory, Princeton University, Princeton, NJ 08544}

\begin{abstract}
\noindent 

Population-III (Pop-III) starformation (SF) is thought to be quenched
when the metallicity of the star-forming gas has reached a critical
level.  At high redshift, when the general intergalactic medium (IGM)
was enriched with metals, the fraction of primordial gas already
collapsed in minihalos (above the Jeans mass but below the mass
corresponding to the efficient atomic cooling threshold for SF) was
significantly larger than the fraction of primordial gas that had
already been involved in Pop-III SF.  We argue that this reservoir of
minihalo gas remained largely in a metal-free state, until these
minihalos merged into large systems (above the hydrogen cooling
threshold) and formed stars.  As a result, the era of Pop-III SF was
significantly prolonged, leading to a total integrated Pop-III SF that
was an order of magnitude larger than expected for an abrupt
transition redshift.  We find that the contribution of Pop-III SF to
the reionization of hydrogen could have been significant until $z\sim
10$ and may have extended to redshifts as low as $z\sim 6$.  
Our modeling allows for {\it gradual} enrichment of the IGM,
feedback from photo-ionization and screening of reionization by minihalos. 
Nevertheless, the extended epoch of Pop-III SF may result in
complex reionization histories, where multi peaks are possible 
over some regions of parameter space.
The relative contribution of Pop-III stars to reionization 
can be quantified and will be tested by three-year WMAP results: 
(1) if Pop-III stars do not contribute to reionization, 
$\tau_{\rm es}\le 0.05-0.06$ and a rapid reionization at $z\sim 6$
are expected with the mean neutral fraction quickly exceeding $50\%$ 
at $z\sim 8$;
(2) if the product of star formation efficiency and escape fraction 
for Pop-III stars is significantly larger than that for Pop-II stars,
then a maximum $\tau_{\rm es}=0.21$ is achievable;
(3) in a perhaps more plausible scenario where 
the product of star formation efficiency and escape fraction 
for Pop-III stars is comparable to that for Pop-II stars,
$\tau_{\rm es}=0.09-0.12$ would be observed, with reionization histories
characterized by an extended ionization plateau from $z=7-12$ where
the mean neutral fraction stays in a narrow range of $0.1-0.3$. 
This result holds regardless of the
redshift where the IGM becomes enriched with metals.  
\end{abstract}


\section{Introduction}

Following cosmological recombination at a redshift $z\sim 10^3$, the
baryonic gas filling up the universe became predominantly neutral.
Given that this gas is known to be mostly ionized today, one arrives
at one of the major questions in current extragalactic astronomy,
namely, when was the cosmic hydrogen re-ionized, and what were the
sources responsible? Recent observations have begun to provide
preliminary answers to this question.  The absorption spectra of SDSS
quasars at $z\sim 6$ indicate that the neutral fraction of hydrogen
increases significantly at $z\ga 6$ (Becker et al. 2001; Fan et
al. 2002; Cen \& McDonald 2002; Pentericci et al.~2002; White et
al.~2003; Wyithe \& Loeb~2004; Mesinger \& Haiman~2004; Fan et
al.~2005).  On the other hand, analysis of the first year data from
the {\em WMAP} satellite suggests a large optical depth to electron
scattering $\tau_{\rm es}=0.17\pm 0.04$, implying that the universe
was significantly reionized at redshifts as high as $\sim 17\pm 5$
(Kogut et al.~2003; Spergel et al.~2003). While this result is not
surprising in view of earlier calculations of the reionization history
by metal-free stars (e.g. Wyithe \& Loeb 2003a; Cen 2003a), it offers
an intriguing {\it empirical} path for answering questions regarding
the sources of reionization.

Several authors have pointed out that the combination of these results
can only be reconciled with an extended or multi-peaked reionization
history (Wyithe \& Loeb~2003a,c; Cen 2003a,b; Chiu, Fan, \& Ostriker
2003; Haiman \& Holder~2003; Somerville et al.~2003; Gnedin 2004). A
common feature of these models is that the first generation of stars
plays a significant role.  The first stars in the universe (Pop-III)
formed out of metal-free gas, relic from the big bang. Simulations of
metal-free star formation indicate that the first stars may have high
masses ($M\ga 100M_\odot$), since gas cooling by molecular hydrogen
(H$_2$) cannot lower the gas temperature below $\sim 200$K (Bromm,
Coppi, \& Larson 2002; Abel, Bryan \& Norman 2002).  Massive,
metal-free stars shine at their Eddington luminosity, $L_{\rm E}
\propto M$, and have roughly constant effective (surface) temperatures
of $\sim 10^5$K and lifetimes of $\sim 3\times 10^6$yr, independent of
their mass (Bromm, Kudritzki, \& Loeb 2001). This implies that the
number of ionizing photons produced per baryon incorporated into these
stars ($\sim 8\times10^4$) was independent of their mass function, and
larger by more than an order of magnitude than that of the observed
Population-II (Pop-II) metal-rich stars. As a result, the emissivity
of ionizing photons may be double peaked, leaving open the possibility
of an early reionization by Pop-III stars, with an epoch of
recombination following enrichment of the IGM above the critical
threshold, before a final reionization by Pop-II stars near
$z\sim6$. Indeed, the observed enrichment of the IGM is consistent
with a substantial contribution from Pop-III stars (Oh, Nollett, Madau
\& Wasserburg 2001). On the other hand, it has been suggested that the
non-monotonic behavior of these models is due to the assumption of a
sharp transition in metallicity (Furlanetto \& Loeb~2005). These
authors find that the various feedback mechanisms lead to an extended
but monotonic reionization history.

In this paper we develop a model which follows the densities of
primordial and enriched gas in collapsed systems, and compute
co-dependent SF, reionization and metal enrichment histories.  While
our model is more detailed than many previous semi-analytic approaches
in several respects, the primary qualitative difference is that
Pop-III SF is allowed to proceed in gas that had already collapsed in
minihalos prior to enrichment of the IGM, but which has not formed
stars, even when the general IGM has been enriched to a high
metallicity level.  We motivate the physics of this conclusion, in
addition to modeling its consequences for the reionization history. We
show that the additional post enrichment Pop-III SF boosts the total
integrated Pop-III SF by an order of magnitude in typical cases.

Throughout the paper we assume cosmological parameters obtained
through fits to {\em WMAP} data (Spergel et al.~2003). These include
density parameter values of $\Omega_{m}=0.27$ in matter,
$\Omega_{b}=0.044$ in baryons, $\Omega_\Lambda=0.73$ in a cosmological
constant, and a Hubble constant of $H_0=71~{\rm km\,s^{-1}\,Mpc^{-1}}$
(or equivalently $h=0.71$). For calculations of the
Press-Schechter~(1974) mass function (with the modification of Sheth
\& Tormen~1999) we assume a primordial power-spectrum with a power-law
index $n=1$ and the fitting formula to the exact transfer function of
Cold Dark Matter, given by Bardeen et al.~(1986).  We adopt an rms
amplitude of $\sigma_8=0.84$ for mass density fluctuations in a sphere
of radius $8h^{-1}$Mpc.

\section{Ionizing radiation from Pop-III and Pop-II Stars}

We make the distinction between the ionizing radiation field due to a
possible early population of zero-metallicity stars, and the metal
enriched stars observed at lower redshifts. It is thought that the
primordial initial mass function favored massive stars (Bromm, Copi \&
Larson~1999, 2001; Abel, Bryan \& Norman~2000; Mackey, Bromm,
Hernquist~2002).  The possible existence of this population is very
important for reionization because the spectrum of these stars would
result in an order of magnitude more ionizing photons per baryon
incorporated into stars (Bromm, Kudritzki, \& Loeb 2001).  The
formation of the very massive stars is suppressed as the material out
of which stars form is enriched with metals.  The fraction of the
ionizing photons produced by metal-free stars depends on several
unknown parameters, including the mixing efficiency of metals, the
environments in which new stars form, and most importantly, the
threshold metallicity above which star formation is transformed from
being dominated by massive stars to a Scalo~(1998) initial mass
function (IMF). The threshold metallicity is believed to be small;
Bromm et al.~(2001) argue for a threshold $\frac{Z_{\rm
thresh}}{Z_\odot}\la 10^{-3}$ of the solar metallicity value.  The
efficiency of mixing of metal enriched outflows from star forming
galaxies to the surrounding IGM is even more uncertain; Scannapieco,
Ferrara \& Madau~(2002) and Ricotti \& Ostriker (2004) find that the
mass weighted mean metallicity can reach values greater than $10^{-3}$
of the solar value at redshifts as high as 20, and note that the
average metallicity scales with star formation efficiency, supernova
rate, and the fraction of supernova energy that is channeled into
outflows.  In a representative region of the IGM, the average
metallicity increases roughly exponentially with redshift as it is
modulated by the exponential growth in the collapse fraction of
baryons at high redshifts. We therefore define a transition redshift
$z_{\rm tran}$ below which metal-rich stars with a Scalo~(1998) IMF
dominate the production rate of ionizing photons. However, because
enrichment will not occur at the same time in all parts of the IGM, we
allow for a spread in the transition redshift. We assume a Gaussian
probability function with variance $\Delta z_{\rm tran}$, so that the
fraction of IGM $(f_{\rm tran})$ that is enriched at redshift $z$ is
given by
\begin{equation}
f_{\rm tran} = \frac{1}{\sqrt{2\pi}(\Delta z)}\int_{-\infty}^z
dz^\prime \exp{\left(\frac{-(z-z_{\rm tran})^2}{2(\Delta
z)^2}\right)}.  
\end{equation}

The effectiveness of a stellar population in ionizing hydrogen in the
IGM can be parameterized in terms of the number of ionizing photons
produced per baryon incorporated into stars ($N$). Throughout the
paper we assume Pop-III stars to be massive ($\ga 100 M_\odot$)
zero-metallicity stars with the generic spectrum calculated by Bromm,
Kudritzki \& Loeb~(2001). The resulting value is $N_{\rm
primord}=84168$. We assume metal enriched stars (1/20th solar
metallicity) to have a Scalo~(1998) mass-function, and use spectral
information from the stellar population model of Leitherer et
al.~(1999)\footnote{Model spectra of star-forming galaxies were
obtained from http://www.stsci.edu/science/starburst99/.}.  This
results in a value of $N_{\rm enrich}\sim4270$.

\section{Enrichment of primordial gas}

Throughout the paper we assume that star-formation is initiated
through atomic hydrogen cooling, and so implicitly assume the rapid
destruction of molecular hydrogen in the IGM by a back-ground of UV
photons; molecular hydrogen had been destroyed long before the
universe, on average, became significantly ionized (an ionized
fraction of $\sim 10^{-3}$) (Haiman, Rees \& Loeb~1997; Oh \&
Haiman~2003).  Therefore, at the time when the IGM started to become
enriched to the critical level for Pop-II SF there was significant gas
already collapsed in minihalos (halos above the cosmological Jeans
mass, but below the minimum mass for SF) that had virial temperatures
which were too low to initiate SF (see \S~\ref{primordev}). If this
gas was preserved in its primordial state until SF was initiated
through formation of a sufficiently massive system, then the epoch of
Pop-III SF could have been increased to redshifts significantly below
$z_{\rm tran}$.

\newpage
\subsection{Enrichment of Virialized Primordial Gas In Minihalos}
\label{haloenrich}

It is often assumed that metal enrichment of the intergalactic medium
(IGM) implies metal enrichment of all gas in the universe (e.g.,
Wyithe \& Loeb 2003a,b; Cen~2003a,b Furlanetto \& Loeb 2005).  We
shall point out that this simple assumption is incorrect, which, as we
will show, may lead to qualitatively different conclusions with regard
to contributions of Pop-III stars to cosmological reionization.  A
critical point to note is that metal enrichment process facilitated by
shockwaves carrying metals generated in stars is a strong function of
the density of the gas that is being enriched.  In particular, we will
make the distinction between metal enrichment of the average IGM and
metal enrichment of gas that is inside already formed halos.

Let us consider metal enrichment of gas in minihalos when a blastwave
of velocity $U$ sweeps through the IGM.  This is equivalent to the
case of a gas cloud moving through a background medium with velocity
$U$.  The gas cloud may be subject to a host of hydrodynamical
instabilities, including Kelvin-Helmholtz and Richtmeyer-Meshkov
instabilities, through which metal mixing occurs.  Several authors
(Murray et al. 1993, M93 hereafter; Klein, McKee, \& Colella 1994;
Dinge 1997; Miniati et al. 1997) have shown that a
non-self-gravitating gas cloud moving at the sound speed of the
background medium gets disrupted on a time scale of the dynamic time
of the cloud.  Self-gravity, in our case by the dominant dark matter
in a minihalo, could stabilize the gas cloud.  M93 studied the case of
self-gravitating, non-cooling gas clouds moving through a background
medium in the context of a two-phase medium.  The gas clouds in our
minihalos are physically identical to the case considered by M93 in
that there is little cooling (due to lack of cooling agents at the low
temperature), and the gas cloud is confined by gravitational potential
well of the dark matter minihalo.  Therefore, we may draw directly
upon the calculations by M93 to quantify the whether or not gas clouds
in minihalos survive enrichment of the IGM.  M93 show both
analytically and by simulations that the stability condition for a gas
cloud moving through a background medium requires the surface gravity
of a cloud to be greater than 
\begin{equation} 
g_{\rm c} \equiv {2\pi U^2\over D R_{\rm cl}},
\end{equation} 
\noindent where $D$ is the
density ratio of the gas cloud to the background gas, $R_{\rm cl}$ is
the radius of the gas cloud.  If we define $\eta$ as 
\begin{equation}
\eta = {g D R_{\rm cl} \over 2\pi U^2}, 
\end{equation} 
\noindent then
we find that
\begin{equation} 
\label{eta} 
\eta(r) = 8.5 ({U\over 10
\mbox{km/s}})^{-2} ({M_{\rm h}\over 10^6 h^{-1}M_\odot})^{2/3}
({1+z\over 15}) ({M_r\over M_{\rm h}})^{-4.7}
\end{equation} 
\noindent where we have assumed that the density slope near the virial 
radius is $-2.4$ (Navarro, Frenk, \& White 1997); $M_{\rm h}$ is the mass of the
minihalo within its virial radius; $M_r$ is the mass with radius $r$,
$z$ is redshift.  M93 show at $\eta=1$ only $2\%$ and $11\%$ of the
gas is lost after $3.2$ and $10$ times the dynamic time of the cloud,
which is about the Hubble time.  However, they also note that even at
$\eta=0.25$, the gas mass loss is still relatively small.

While $U$ might be large in the immediate vicinity of a
shockwave-producing galaxy, one does not expect $U$ to be large at
large distances.  Mori, Ferrara, \& Madau (2002) show, in simulations
of propagation of supernova blastwaves from $10^8h^{-1}M_\odot$
galaxies at $z=9$, that after more than a hundred million years the
relative filling factor for regions being swept by shocks of
velocities larger than $U=(10,30,100)\mbox{km/s}$ is roughly $(100\%,
35\%, 10\%)$.  We expect the velocities to be still smaller at the
higher redshifts of concern here, due to enhanced cooling and larger
Hubble velocity.  We see from equation~(\ref{eta}) that if $U\le 29
\mbox{km/s}$ $100\%$ of the gas within virialized regions of halos
more massive than $10^6h^{-1}M_\odot$ will have $\eta\ge 1$ and be
relatively un-affected by the shockwaves.  This fraction is reduced to
$60\%$ and $30\%$ for $U=100\mbox{km/s}$ and $U=500\mbox{km/s}$,
respectively.  For lower mass mini-halos ($M_{\rm
h}\sim10^{4.5}M_\odot$) close to the Jeans mass (\S~\ref{massscales}),
we find $\eta\sim1$ for a smaller velocity of
$U\sim10\mbox{km/s}$. However $\eta(r)$ is a high power of the
fraction of gas in the halo that is subject to instabilities. Thus
even for halos of mass $M_{\rm h}\sim10^{4.5}M_\odot$, and winds of
velocity $U\sim30$km/s and $U\sim100$km/s we find values of ${M_r\over
M_{\rm h}}\sim0.6$ and ${M_r\over M_{\rm h}}\sim0.35$ yield
$\eta \sim1$. Combining these results, we expect that most of the gas
already virialized with minihalos will be largely unaffected by
metal-carrying blastwaves and remain metal-free until its first
episode of starformation.

In addition to direct enrichment, there is the possibility that
virialized primordial gas in mini-halos could be enriched during the
process of mergers with other mini-halos containing enriched gas. We
do not expect this to be the case for the reasons outlined below.
Following the transition redshift, if the blastwaves that carry metals
sweep through the IGM with a velocity of order of $\sim 10$~km/s, the
IGM may be collisionally ionized.  The collisionally heated IGM
subsequently cools via the Compton cooling process (by the cosmic
microwave background), and at the same time recombines in the absence
ionizing sources.  We find that collisionally ionized gas would be
Compton-cooled and saturate at a temperature of $400-500$K (depending
on the initial temperature) after about a Hubble time at $z=15$; at
1/4 of the Hubble time $T\ge 600$~K.  Hence, IGM that is heated at the
redshift of interest will not be able to cool below $\sim 400$~K by
radiative processes.  Subsequent cooling is primarily due to adiabatic
expansion, which could achieve a temperature reduction by a factor of
a few for the redshift range of interest.  For a IGM of $T\sim 400$~K
at $z=15$ the Jeans mass is $\sim 4\times 10^6M_\odot$, which is close
to the threshold mass for atomic cooling (see equation 7 below).
Therefore, metal-enriched and heated IGM is likely to be prevented
from further accreting onto all minihalos, new and old.  Thus, gas in
collapsed minihalos will not be contaminated by enriched IGM gas and
will remain metal-free, and newly formed minihalos will be gas-free.

To be extremely conservative, let us consider the case where the
metal-enriched and heated IGM is assumed to have cooled to the
temperature of the cosmic microwave background.  Gas from the IGM can
enter mini-halos with $M\ga 10^5-10^6$M$_\odot$ either through mergers
of smaller collapsed objects, or through accretion from the IGM. In
the former case, the cold gas content is primordial because accretion
from the IGM at the temperature of the cosmic microwave background is
suppressed in low mass mini-halos. In the latter case we argue that
newly accreted gas into more massive mini-halos is unlikely to be as
dense as the gas already at the virial radius. This may be seen as
follows. An estimate of the non-linear overdensity of baryons inside a
virialized object is 
\begin{equation} 
\label{baryonod} 
\delta_{\rm
b}=\frac{\rho_{\rm b}}{\bar{\rho}_{\rm b}}-1 =
\left(1+\frac{6}{5}\frac{T_{\rm vir}}{\bar{T}}\right)^\frac{3}{2}-1
\approx \frac{9}{5}\frac{T_{\rm vir}}{\bar{T}}, 
\end{equation} 
where $\rho_{\rm b}$ is the density of baryons inside the virialized
object of temperature $T_{\rm vir}$, and $\bar{\rho}_{\rm b}$ and
$\bar{T}$ are the background baryon density and temperature
respectively (Barkana \& Loeb~2001). At redshifts below $z\sim130$,
and in the absence of a heating source the temperature of an
adiabatically cooling IGM is $T_{\rm g}\sim
6.8\left(\frac{1+z}{20}\right)^2$K (Peebles 1993). This temperature is
substantially lower than the temperature of the cosmic microwave
background, $T_{\rm CMB}\sim 54.6\left(\frac{1+z}{20}\right)$. Thus
primordial gas, which has not been subjected to heating accretes into
a dark-matter halo with an initial temperature that is substantially
lower than does primordial gas which has been heated by the enriching
blast-wave before being Compton cooled to the temperature of the
cosmic microwave background.  From equation~(\ref{baryonod}) we
therefore find that the ratio of overdensities of primordial to
enriched gas following virialization is
$\sim8\left(\frac{1+z}{20}\right)^{-1}$. This result suggests that in
the absence of cooling, enriched gas would be prevented from
significantly penetrating the minihalo, so that mixing is expected to
be inefficient. As a result primordial gas is not expected to become
contaminated during hierarchical growth of mini-halos.

Upon merging to form a system large enough to initiate SF, does the
primordial gas in mini-halos form stars before becoming enriched by
other SF in the galaxy?  This is expected to happen to some extent.
But ``other" SF that occurred earlier would have been Pop-III SF.
Therefore, the question is not whether Pop-III SF will occur or not,
but rather what the efficiency of Pop-III formation will be.  We do
not have a quantitative answer to this at present.  It seems plausible
to expect that the first SF would occur in the largest mass clump in
the galaxy.  If the largest mass clump makes up a significant fraction
of the total mass of the galaxy, then the overall SF formation
efficiency may be largely determined by the SF efficiency within that
largest mass clump.  We will therefore assume, following conventional
wisdom, that Pop-III SF proceeds with efficiency $\eta=0.1$.

In summary, the above discussion leads us to explore the effect of the
following hypothesis on the Pop-III SF history, and on the
reionization history of cosmic hydrogen. 
$i)$ We assume that virialized primordial gas within mini-halos is not enriched by
super-galactic winds. $ii)$ Virialized primordial gas is not enriched
following accretion of enriched gas from the IGM, or during the
mergers of mini-halos. $iii)$ As a result, primordial virialized gas
remains in its primordial state until such time as its host halo grows
to be massive enough to initiate a burst of Pop-III SF.

\section{Evolution of the density of gas in halos}

In order to compute the effect of IGM enrichment on the SF history, we
need to follow the density of enriched and primordial gas through the
hierarchical merging of halos. This section introduces a method for
following this evolution.  During hierarchical galaxy formation, an
average property of the galaxy population, such as the density of
galaxies evolves in redshift due to newly collapsing halos, mergers of
halos and accretion. We begin by discussing the critical mass-scales
in galaxy formation. We then discuss calculation of the rate of newly
collapsing halos, before describing how this quantity, along with the
merger rates of galaxies can be used as source and sink terms in
differential equations that describe the average evolution of the gas
content of the galaxy population.

\subsection{Critical Mass Scales}
\label{massscales}

There are four critical mass scales affecting galaxy formation. The
first is the Jeans mass, which corresponds to the smallest mass halo
into which gas can accrete from the IGM. By considering the response
of a baryonic overdensity to the potential well of a dark-matter halo
one can find the mass scale at which the baryonic overdensity reaches
100. We take this mass scale to represent the Jeans mass (Barkana \&
Loeb~2001) 
\begin{equation} 
\label{jeans} 
M_{\rm J}=5\times10^3\left(\frac{\Omega_m
h^2}{0.15}\right)^{-\frac{1}{2}}\left(\frac{\Omega_b
h^2}{0.022}\right)^{-\frac{3}{5}}\left(\frac{1+z}{10}\right)^{\frac{3}{2}}M_\odot.
\end{equation} 

The second mass scale corresponds to the virial
temperature above which gas that has accreted into a halo can cool
efficiently. We assume this scale to be determined by atomic cooling,
and so to correspond to a virial temperature below $T_{\rm
min}\sim2\times10^4$K. The corresponding mass is therefore
\begin{equation} 
M_{\rm min} = 10^8h^{-1}\left(\frac{T_{\rm
min}}{1.98\times10^{4}K}\right)^{\frac{3}{2}}\left(\frac{0.6}{\mu_P}\right)^\frac{3}{2}
\left(\frac{\Omega_{\rm
m}}{\Omega_z}\frac{\Delta_c}{18\pi^2}\right)^{-\frac{1}{2}}\left(\frac{1+z}{10}\right)^{-\frac{3}{2}},
\end{equation} 
where $\Omega_m^z\equiv[1+(\Omega_\Lambda/\Omega_m)(1+z)^{-3}]^{-1}$,
$\Delta_c=18\pi^2+82d-39d^2$ and $d=\Omega_m^z-1$ (see Barkana \&
Loeb~2001 for more details). The third mass scale becomes important
following the reionization of a region of IGM at which time it is
heated to $\sim10^4$K. The Jeans mass is increased by several orders
of magnitude, and numerical simulations find that gas infall is
suppressed in halos with $T_{\rm vir}\la2.5\times10^5$. There is some
disagreement as to the exact value of the halo circular velocity below
which gas infall is completely suppressed (e.g.  Thoul \&
Weinberg~1996; Kitayama \& Ikeuchi~2000; Quinn, Katz \&
Efstathiou~1996, Weinberg, Hernquist \& Katz~1997; Navarro \&
Steinmetz~1997; Dijkstra, Haiman, Rees \& Weinberg~2004). In what
follows we assume a post-reionization minimum temperature of $T_{\rm
reion}\sim2.5\times10^5$, and note that the assumption of a smaller
temperature will not qualitatively effect our conclusions. Thus we
have 
\begin{equation} 
M_{\rm reion} = 4.5\times 10^9h^{-1}\left(\frac{T_{\rm
reion}}{2.5\times10^{5}K}\right)^{\frac{3}{2}}\left(\frac{0.6}{\mu_P}\right)^\frac{3}{2}
\left(\frac{\Omega_{\rm
m}}{\Omega_z}\frac{\Delta_c}{18\pi^2}\right)^{-\frac{1}{2}}\left(\frac{1+z}{10}\right)^{-\frac{3}{2}}.
\end{equation}

Finally, the fourth characteristic mass is the non-linear mass-scale
($M_{\rm nl}$), which corresponds to the co-moving length scale
$R_{\rm nl}=(\frac{3M_{\rm nl}}{4\pi\rho_{\rm m}})^{1/3}$ over which
typical fluctuations (i.e. 1-sigma fluctuations) in the linear density
field have a value of $\sigma(M_{\rm nl})=\delta_{\rm
crit}/D(z)$. Here $\rho_{\rm m}$ is the co-moving density of matter,
$\delta_{\rm crit}$ is the linearly extrapolated overdensity at the
time of collapse for a dark matter halo, and $D(z)$ is the growth
factor at redshift $z$. For the cosmology employed in this paper, and
at redshifts $z\ga1$, a good fit to the dependence of the non-linear
mass scale on redshift is given by
\begin{equation}
\label{NL}
M_{\rm nl}(z) = 10^{13.58}(1+z)^{-1.39}\exp{\left[\left(\frac{z}{0.18}\right)^{0.79}\right]}.
\end{equation}
From equation~(\ref{NL}) we see that $M_{\rm nl}$ moves through the
Jeans mass at $z\sim8$, and the mass corresponding to the hydrogen
cooling threshold ($M_{\rm min}$) at $z\sim4$. The non-linear
mass-scale corresponds to the typical mass of newly forming
galaxies. Equation~(\ref{NL}) therefore encapsulates the reason why
reionization is expected to occur at redshifts between $z\sim10$ and
$z\sim5$. Reionization is a self-limiting process because the
non-linear mass-scale does not reach $M_{\rm reion}$ until $z\sim2$.

\subsection{Primordial Gas In Collapsed Systems At $z_{\rm tran}$}

\begin{figure*}[hptb]
\epsscale{0.4}
\plotone{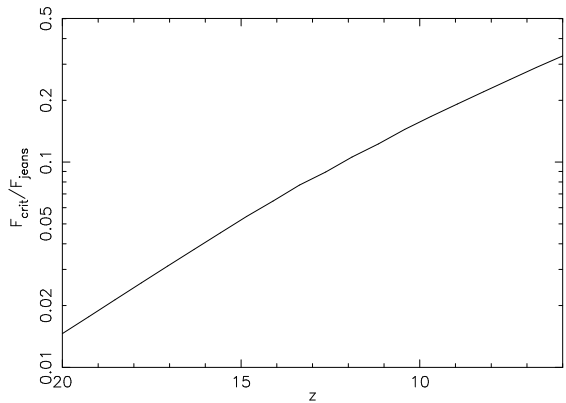}
\caption{\label{fig1} The ratio of collapsed fractions of gas in galaxies with masses above the Jeans mass, and above $M_{\rm min}$ respectively.}
\end{figure*}

Prior to the enrichment of the IGM with metals produced by the first
stars the gas content of collapsed systems below the critical
threshold for SF (minihalos) was metal-free. Systems above the
critical threshold had an initial burst of metal-free SF, but then
presumably became enriched internally so that subsequent SF was metal
enriched. Following enrichment of the IGM to a metallicity above a
critical value, gas that was newly accreted from the IGM was enriched,
and so resulted in Pop-II SF. However, the metal-free gas in minihalos
would not have been enriched and would have moved through mergers into
larger systems. When this metal-free gas became part of a galaxy above
the SF threshold, further Pop-III SF would have been possible, even
significantly after the IGM became enriched. Indeed, if one calculates
the collapsed fraction of gas in galaxies at $z=15$ with masses above
the Jeans mass, to the collapsed fraction above the mass corresponding
to the minimum for SF ($M_{\rm min}$) one finds that the collapsed
fraction differs by a factor of 20!  Figure~\ref{fig1} shows the ratio
of collapsed fractions above $M_{\rm min}$ and the Jeans mass as a
function of redshift. The majority of metal-free gas that has accreted
into collapsed systems at the time of enrichment has not yet formed
stars. As a result the time of enrichment may not correspond to the
time when Pop-III SF ends. Indeed the fraction of the Pop-III SF
during the history of the universe that has taken place by the
transition redshift could be as low as the ratio shown in
Figure~\ref{fig1}. We therefore expect most Pop-III SF to take place
at redshifts below $z_{\rm tran}$.

%FIGURE 1

\subsection{The Rate of Newly Collapsing Dark-Matter Halos}

To follow the evolution of primordial gas in a hierarchical cosmology
we first need to compute the collapse rate of new halos.  Let
$\frac{dn_{\rm ps}}{dM}(z)$ be the Press-Schechter~(1974) mass
function (number of halos with mass between $M$ and $M+dM$ per
co-moving Mpc$^3$) of dark-matter halos at redshift $z$. The rate of
change of the density of dark matter halos between $M$ and $M+dM$ is
therefore $\frac{d^2n_{\rm ps}}{dM_1dz}$. Let us also define (Lacey \&
Cole~1993) $\frac{d^2N_{\rm mrg}}{d\Delta Mdt}\left.\right|_{M}$ the
number of mergers per unit time of halos of mass $\Delta M$ with halos
of mass $M$ (forming new halos of mass $M_1=M+\Delta M$) at redshift
$z$. We find the component of the change in density with redshift
of halos having masses between $M$ and $M+dM$ that is not due to
mergers.  This is given by the overall change in the density of halos
with mass $M$, minus the change in density due to merger of halos to
form a new halo of mass $M$, plus the change in density of halos of
mass $M$ that merge with other halos. The collapse rate of new halos
may be written
\begin{equation} 
\label{col} 
\frac{d^2n_{\rm col}}{dM_1dz} =
\frac{d^2n_{\rm ps}}{dM_1dz}-\int_{0}^{\frac{M_1}{2}}d\Delta
M\left.\frac{d^2N_{\rm mrg}}{d\Delta Mdt}\right|_{M_1-\Delta
M}\frac{dt}{dz}\frac{dn_{\rm ps}}{d(M_1-\Delta M)} +
\int_{0}^{\infty}d\Delta M\left.\frac{d^2N_{\rm mrg}}{d\Delta
Mdt}\right|_{M_1}\frac{dt}{dz}\frac{dn_{\rm ps}}{dM_1}.
\end{equation} 
The value of $\frac{d^2n_{\rm col}}{dM_1dz}$ equals the net change in
density due to newly collapsed halos of mass $M$.

\subsection{The Evolution Of The Primordial Gas Content Of Galaxies}
\label{primordev}

Our next step is to find the evolution of the density of metal-free
gas per halo mass $d\rho_{\rm primord}/dM$. {\it Note that this
density refers to a density of gas that has not been in a galaxy at a
time when it underwent a burst of SF}. We assume that if a halo forms
with a mass $M_1$ larger than the Jeans mass $M_{\rm J}$, or if the
mass of a halo grows through a merger to become larger than $M_{\rm
J}$, then enough gas accretes inside the virial radius so that the
mass of gas inside the halo becomes $f_{\rm b}M$, where $f_{\rm b}$ is
the fraction of the mass-density in baryons. Therefore, if the halo
mass is larger than $M_{\rm J}$, then the total density of gas inside
halos in a small mass-range between $M_1$ and $M_1+\delta M$ is
\begin{equation} 
\frac{d\rho_{\rm primord}}{dM_1}\delta M=f_{\rm
b}M\frac{dn_{\rm ps}}{dM_1}\delta M.  
\end{equation} 
Our aim is to compute the effect of metal enrichment on the density of
cold, metal free gas inside galaxies of mass $M_1$.

Prior to enrichment of the IGM, the gas in galaxies where SF has not
occurred remains metal-free.  Following enrichment, newly accreted gas
is metal enriched. In neutral regions the gas inside halos below
$M_{\rm min}$ is not photo-evaporated, and so, at least initially some
halos with masses below $M_{\rm min}$ will contain metal-free
gas. This gas will move into larger halos during subsequent
mergers. We can calculate the density of metal-free gas that remains
in these low mass halos following enrichment of the IGM. If two halos
$M$ and $\Delta M$ merge to form a new larger halo $M_1$ with a mass
smaller than $M_{\rm min}$, then the average mass of metal-free gas in
the new halo will equal the sum of the average masses of metal-free
gas in each of the initial halos $(\frac{d\rho_{\rm
primord}}{dM}/\frac{dn_{\rm ps}}{dM} + \frac{d\rho_{\rm
primord}}{d\Delta M}/\frac{dn_{\rm ps}}{d\Delta M})$.

We can therefore write down the evolution of the density of primordial
gas which we denote $\frac{d\rho_{\rm primord}}{dM_1}$ ({\it recall
that this density refers to a density of gas that has not been in a
galaxy at a time when it underwent a burst of SF}) 
\begin{eqnarray}
\label{gasev} 
\nonumber \frac{d^2\rho_{\rm primord}}{dM_1dz} &=&
\left(1-f_{\rm tran}(z)\right) \times \left[\frac{d^2n_{\rm
col}}{dM_1dz}f_{\rm b}M_1+\int_{0}^{\frac{M_1}{2}}d\Delta
M\left.\frac{d^2N_{\rm mrg}}{d\Delta Mdt}\right|_{M_1-\Delta
M}\frac{dt}{dz}\frac{dn_{\rm ps}}{d(M_1-\Delta M)}f_{\rm
b}M_1\right]\\ 
\nonumber &+& f_{\rm tran}(z) \times
\int_{0}^{\frac{M_1}{2}}d\Delta M\left.\frac{d^2N_{\rm mrg}}{d\Delta
Mdt}\right|_{M_1-\Delta M}\frac{dt}{dz}\frac{dn_{\rm ps}}{d(M_1-\Delta
M)}f_{\rm b}\left(\frac{\frac{d\rho_{\rm primord}}{dM}}{\frac{dn_{\rm
ps}}{dM}} + \frac{\frac{d\rho_{\rm primord}}{d\Delta M}}{\frac{dn_{\rm
ps}}{d\Delta M}}\right)\\ 
\nonumber &-& \int_{0}^{\infty}d\Delta
M\left.\frac{d^2N_{\rm mrg}}{d\Delta
Mdt}\right|_{M_1}\frac{dt}{dz}\frac{d\rho_{\rm primord}}{dM_1}
\hspace{23mm} \mbox{where}\hspace{5mm}M_{\rm J}<M_1<M_{\rm min}\\
\frac{d^2\rho_{\rm primord}}{dM_1dz} &=& 0\hspace{73mm}
\mbox{otherwise.}  
\end{eqnarray} 
The first equation of (\ref{gasev}) is valid in neutral regions, while
once the region become ionized any remaining primordial gas is
photo-evaporated out of the minihalos on the halo dynamical timescale
($\sim0.1H^{-1}$), where it then becomes enriched, so that
$\frac{d^2\rho_{\rm primord}}{dM_1dz}=0$. The first line contains two
terms that relate to regions of IGM that have not yet been
enriched. The first term corresponds to a source of collapsing halos
into which metal-free gas accretes. The second term corresponds to a
source of gas accumulation in halos above the Jeans mass. This
accumulation is the sum of gas that was already in halos if the
progenitor was above the Jeans mass, and gas newly accreted from the
IGM if it was not. The third term (second line) corresponds to regions
with enriched IGM and describes the movement of primordial gas from
small halos to large halos during mergers. The fourth term describes
the loss of gas density in halos of mass $M$ that results from the
merger of those halos to form larger systems. The primordial gas
density in halos smaller than the Jeans mass or larger than $M_{\rm
min}$ is zero. The upper left panels of figures~\ref{fig2} and
\ref{fig3} show the evolution of the primordial gas density with
redshift assuming $z_{\rm tran}=17$ with $\Delta z_{\rm tran}=1.25$,
and $z_{\rm tran}=22$ with $\Delta z_{\rm tran}=2.5$. The lines show
the evolution density at constant halo mass. Prior to $z_{\rm tran}$
these follow the expectation from the Press-Schechter mass function.

\subsection{The Evolution Of The Metal-Enriched Gas Content Of Galaxies}

We compute the density of enriched gas inside halos of mass $M_1$. In
regions of the IGM that have not yet been enriched, newly accreted gas
is primordial. Conversely in enriched regions of the IGM, newly
accreted gas is metal enriched.  Prior to the reionization of a
region, gas can cool inside a halo whose virial temperature is larger
than $T_{\rm min}\sim2\times10^4$K. However, following reionization of
a region, infall is suppressed in halos whose virial temperature is
smaller than $T_{\rm reion}\sim2.5\times10^5$K. In neutral regions the
enriched gas inside halos below $M_{\rm min}$ is not photo-evaporated
so that, at least initially, some halos with masses below $M_{\rm
min}$ will contain enriched gas that has not undergone SF. This gas
will move into larger halos during subsequent mergers. Using a method
similar to that of the preceding section, we can calculate the density
of enriched gas that remains in these low mass halos following
reionization. Note here that we assume enriched gas can accrete into
halos of all masses in excess of $M_{\rm J}$. Conversely, in
\S~\ref{haloenrich} we argued that the IGM will remain at a
temperature of $\sim500$K following enrichment by a blastwave,
implying that accretion of enriched gas is suppressed for mini-halos
with masses smaller than $\sim5\times10^6$M$_\odot$. However this
inconsistency will not affect our estimate of Pop-II SF since
$\sim5\times10^6$M$_\odot<M_{\rm min}$.

%FIGURE 2 
\begin{figure*}[t] 
\epsscale{.8} 
\plotone{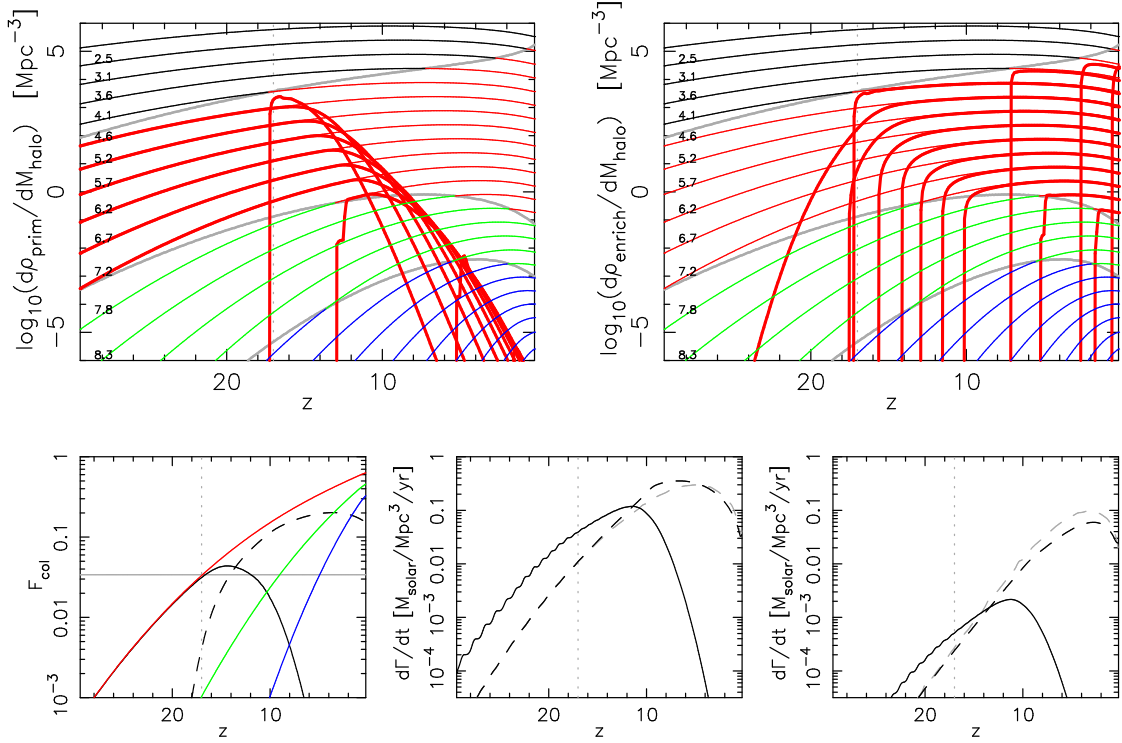}
\caption{\label{fig2} {\em Upper panels:} The evolution of the density
of gas {\em that has not been involved in a starburst}, within dark
matter halos as a function of redshift. The redshift where the IGM
became metal enriched was taken to be $z_{\rm tran}=17$, with a
variance of $\Delta z_{\rm tran}=1.25$. This epoch is marked by the
vertical dotted line. The thin curves show the density of gas
corresponding to the maximum possible density, i.e. $\frac{dn_{\rm
gas,enrich}}{dM}=f_{\rm b}M\frac{dn_{\rm ps}}{dM}$ where $f_{\rm
b}=\Omega_{\rm m}/\Omega_{\rm b}$. The lines are labeled by the
logarithms (base 10) of the corresponding masses. The grey curves show
the corresponding maximum gas density for the $M_{\rm J}$, $M_{\rm
min}$ and $M_{\rm reion}$. The thick lines show the evolution of the density
of gas that has not yet undergone a star-burst. The upper left and
upper right panels show the evolution of primordial gas $\frac{dn_{\rm
primord}}{dM}$, and enriched gas $\frac{dn_{\rm enrich}}{dM}$
respectively. {\em Lower panels:} Collapsed fractions and SF. The
lower left panel shows the collapsed fraction of primordial gas (solid
line) and enriched gas that has not yet undergone a starburst (dashed
lines). For comparison we show the collapsed fractions of dark-matter
halos above $M_{\rm J}$, $M_{\rm min}$ and
$M_{\rm reion}$ (left to right). The lower-middle and lower-right panels show
the SF rates in neutral regions for Pop-III stars (solid curves) and
for Pop-II stars (dashed curves), as well as for Pop-II stars in
reionized regions (dashed grey curves). The lower-middle and right-hand
panels show SF in cases-A and B respectively (see \S 5).}
\end{figure*}

%FIGURE 3
\begin{figure*}[t]
\epsscale{.8}
\plotone{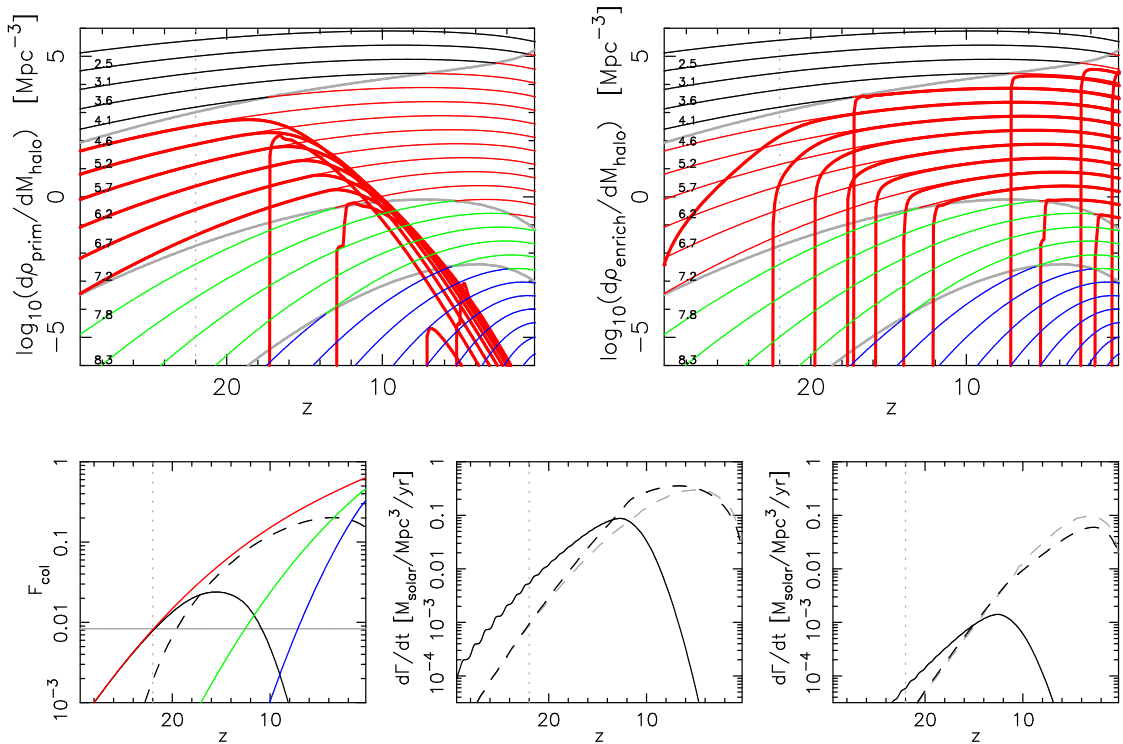}
\caption{\label{fig3} As per Figure~\ref{fig2}, but with $z_{\rm tran}=22$ and $\Delta z_{\rm tran}=2.5$.}
\end{figure*}

We can therefore write down the evolution of the density of enriched
gas which we denote $\frac{d\rho_{\rm enrich}}{dM_1}$ ({\it note that
this density refers to a density of gas that has not been in a galaxy
at a time when it underwent a burst of SF}).  Photo-ionization leads
to $\frac{d\rho_{\rm enrich}}{dM_1}=0$ in ionized regions of the
IGM. In neutral regions we have 
\begin{eqnarray} 
\label{gasev_enrich}
\nonumber 
\frac{d^2\rho_{\rm enrich}}{dM_1dz} &=& f_{\rm
tran}\times\left[\frac{d^2n_{\rm col}}{dM_1dz}f_{\rm b}M_1 +
\int_{0}^{\frac{M_1}{2}}d\Delta M \left.\frac{d^2N_{\rm mrg}}{d\Delta
Mdt}\right|_{M_1-\Delta M}\frac{dt}{dz}\frac{dn_{\rm ps}}{d(M_1-\Delta
M)}\right. \\ 
\nonumber && \hspace{20mm}\left.\times\{\Theta(M_{\rm
J}-\Delta M)f_{\rm b}\Delta M + \Theta(M_{\rm J}-[M_1-\Delta M])f_{\rm
b}[M_1-\Delta M]\}\frac{\frac{}{}}{\frac{}{}}\right]\\ 
\nonumber &+&
\int_{0}^{\frac{M_1}{2}}d\Delta M\left.\frac{d^2N_{\rm mrg}}{d\Delta
Mdt}\right|_{M_1-\Delta M}\frac{dt}{dz}\frac{dn_{\rm ps}}{d(M_1-\Delta
M)}f_{\rm b}\left(\frac{\frac{d\rho_{\rm enrich}}{dM}}{\frac{dn_{\rm
ps}}{dM}} + \frac{\frac{d\rho_{\rm primord}}{d\Delta M}}{\frac{dn_{\rm
ps}}{d\Delta M}}\right)\\ 
\nonumber &-& \int_{0}^{\infty}d\Delta
M\left.\frac{d^2N_{\rm mrg}}{d\Delta
Mdt}\right|_{M_1}\frac{dt}{dz}\frac{d\rho_{\rm enrich}}{dM_1}
\hspace{23mm} \mbox{where}\hspace{5mm}M_{\rm J}<M_1<M_{\rm min}\\
\frac{d^2\rho_{\rm enrich}}{dM_1dz} &=& 0\hspace{73mm}
\mbox{otherwise,} 
\end{eqnarray} 
where $\Theta$ is the Heaviside step function.  The first two terms
(first line) only contribute in enriched regions of the IGM. The first
term corresponds to a source of collapsing halos in which metal
enriched gas accretes from the IGM. The second term corresponds to a
source term of gas accumulation due to progenitor halos which are
below the Jeans mass. These progenitor halos result in accretion of
enriched gas from the IGM.  The third term describes the movement of
enriched gas from small halos to large halos during mergers. The
fourth term describes the loss of gas density in halos of mass $M$
that results from the merger of those halos to form larger
systems. The enriched gas density in halos smaller than the Jeans mass
or larger than $M_{\rm crit}$ is zero. The upper right panels of
figures~\ref{fig2} and \ref{fig3} show the evolution of the enriched
gas density with redshift assuming $z_{\rm tran}=17$ with $\Delta
z_{\rm tran}=1.25$, and $z_{\rm tran}=22$ with $\Delta z_{\rm
tran}=2.5$.

\subsection{The Collapsed Fraction Of Gas Inside Galaxies}

The fraction of gas in the universe that has collapsed inside galaxies
at redshift $z$ but not undergone a starburst is 
\begin{equation}
F_{\rm primord}(z) = \int_0^\infty dM \frac{1}{f_{\rm b}\Omega_{\rm
m}\rho_{\rm c}}\frac{d\rho_{\rm primord}}{dM}(z) 
\end{equation} 
for primordial gas, and 
\begin{equation} 
F_{\rm enrich}(z) = \int_0^\infty dM \frac{1}{f_{\rm b}\Omega_{\rm
m}\rho_{\rm c}}\frac{d\rho_{\rm enrich}}{dM}(z).  
\end{equation} 
for enriched gas. These are plotted in the lower left panels of
Figure~\ref{fig2} and \ref{fig3} assuming $z_{\rm tran}=17$ with
$\Delta z_{\rm tran}=1.25$, and $z_{\rm tran}=22$ with $\Delta z_{\rm
tran}=2.5$ respectively. For comparison we also show the collapsed
fractions ($F_{\rm col}$) above $M_{\rm J}$, $M_{\rm min}$ and $M_{\rm
reion}$ (left to right respectively). We see that prior to the
transition redshift, nearly all gas inside collapsed systems is
primordial. After the transition redshift this collapsed primordial
gas is replaced with enriched gas. However, the collapsed fraction of
primordial gas does not peak until $z\sim 14$ and $z\sim16$ assuming
$z_{\rm tran}=17$ and $z_{\rm tran}=22$ respectively. The enriched gas
comes to dominate the collapsed gas fraction at around the same
time. Note that $F_{\rm col}(M_{\rm J}) = F_{\rm primord}+F_{\rm
enrich}+F_{\rm col}(M_{\rm min})$.

\section{The Starformation History}

We model SF to occur following the collapse of baryons into a dark
matter halo larger than $M_{\rm min}$, or following a major merger
with progenitors larger than $M_{\rm min}$. Halos that merge to form a
system greater than $M_{\rm min}$ for the first time are able to
produce Pop-II or Pop-III stars. The evolution of gas densities
derived in the previous section may be used to compute the relative
quantities of each.  Barkana \& Loeb~(2000) suggested a model for the
SFR based on the merger history of halos. In their model, gas that has
not previously cooled and undergone an episode of SF in a galaxy forms
stars with efficiency $\eta$ as it cools following a merger that forms
a galaxy larger than $M_{\rm crit}$. In addition, gas in a galaxy of
mass $M>M_{\rm min}$ undergoes an additional starburst if the galaxy
merges with a second galaxy whose mass is larger than $M/2$. Barkana
\& Loeb~(2000) assume that the cold gas in a galaxy has undergone one
previous starburst, so that the remaining gas mass is reduced by a
factor $(1-\eta)$.

Following Wyithe \& Loeb~(2003a) we consider two prescriptions for SF
which bracket the expected scenario of supernova feedback in low mass
galaxies. The scenario described above is referred to as Case-A.  We
also consider a second evolution for the stellar ionizing field,
denoted hereafter as case B. Analysis of a large sample of local
galaxies shows that the ratio $\epsilon=M_\star/M_{\rm halo}$ (where $M_\star$
and $M_{\rm halo}$ are the total stellar and dark matter halo masses
respectively) scales as $\epsilon\propto M_{\rm halo}^{2/3}$ for
$M_{\star}<3\times10^{10}M_{\odot}$, but is constant for larger
stellar masses (Kauffmann et al.~2003). Since SF is thought to be regulated by
supernova feedback (Dekel \& Silk 1986), the important quantity is the
depth of the galactic potential well, or equivalently the halo
circular velocity.  Using the stellar mass Tully-Fisher relation of
Bell \& De Jong~(2001), we find the threshold circular velocity
$v_\star=176~{\rm km~s^{-1}}$ that at $z=0$ corresponds to a stellar
mass of $3\times10^{10}M_{\odot}$. In this case we define $\eta$ as
the SF efficiency in galaxies with circular velocities larger than
$v_\star$. The SF efficiency in smaller galaxies is multiplied by a
factor $\epsilon$, where $\epsilon=1$ for $M>M_{\rm halo}^\star$ and
$\epsilon=\left({M}/{M_{\rm halo}^\star}\right)^{2/3}$ for $M<M_{\rm
halo}^\star$. The expressions for the SF rate in the following
sub-sections have SF efficiencies written as the product
$\epsilon\eta$. Note that while in Case-B $\epsilon$ shows the mass
dependence described above, $\epsilon=1$ for all masses in Case~A.

\subsection{The Pop-III SF Rate}

We assume that there is a burst of Pop-III SF of mass $\epsilon\eta
f_{\rm b} M$ (where $\epsilon\eta$ is the SF efficiency and we assume
$f_{\rm b}=\Omega_b/\Omega_m$) whenever there is a merger of a halo
with mass $M<M_{\rm J}$ to form a new halo with mass $M_1>M_{\rm
J}$. These starbursts comprise all of the Pop-III SF, since once gas
has been involved in a starburst the gas becomes enriched so that a
subsequent shock due to a major merger results in Pop-II SF. Following
reionization of a region of IGM, all gas in halos below $M_{\rm min}$
is photo-evaporated over a timescale $\sim0.1H^{-1}$ where it becomes
enriched, and hence subsequent SF there is assumed to be Pop-II. The
Pop-III SF rate density in neutral regions is therefore given by
\begin{eqnarray} 
\nonumber 
\frac{d\Gamma_{\rm primord}}{dt}(z)&=&
\int_{M_{\rm min}/2}^{\infty} dM\int_{\max(0,M_{\rm
min}-M)}^{\min(M,M_{\rm min})}d\Delta M \left.\frac{d^2N_{\rm
mrg}}{d\Delta Mdt}\right|_{M}\frac{dt}{dz}\frac{dn_{\rm ps}}{dM}\\
&&\hspace{30mm}\times\left(\left[\Theta(M_{\rm min}-M)f_{\rm
b}\epsilon\eta\frac{\frac{d\rho_{\rm primord}}{dM}}{\frac{dn_{\rm
ps}}{dM}}\right] + \left[f_{\rm b}\epsilon\eta\frac{\frac{d\rho_{\rm
primord}}{d\Delta M}}{\frac{dn_{\rm ps}}{d\Delta M}}\right]\right)
\end{eqnarray} 
The Pop-III SF rate in a neutral IGM is shown in the
lower middle (Case-A) and lower right (Case-B) panels of
figures~\ref{fig2} and \ref{fig3} (solid lines) assuming $z_{\rm
tran}=17$ with $\Delta z_{\rm tran}=1.25$, and $z_{\rm tran}=22$ with
$\Delta z_{\rm tran}=2.5$ respectively. The corresponding integrated
Pop-III SF densities follow from 
\begin{equation} 
\Gamma_{\rm
primord}(>z) = \int_\infty^z dz \frac{dt}{dz}\frac{d\Gamma_{\rm
primord}}{dt}(z).  
\end{equation}

As noted in Figure~\ref{fig1}, the collapsed mass of primordial gas at
$z_{\rm tran}$ far exceeds the mass of gas involved in SF up until
that point. Moreover, we see that in the absence of reionization,
Pop-III SF can persist long after the enrichment of the IGM.  Indeed,
Figures~\ref{fig2} and \ref{fig3} show that if enrichment occurred at
$z_{\rm tran}= 17-22$ then Pop-III SF would not peak until redshift
$z\sim12$, and would not cease until $z\sim4$. This result is quite
insensitive to the exact value $z_{\rm tran}$ and to the mode of SF.

\subsection{The Pop-II SF Rate}

In the model of SF employed, gas that has not previously cooled and
undergone an episode of SF in a galaxy forms stars with efficiency
$\epsilon\eta$ as it cools following a merger to form a galaxy
larger than $M_{\rm crit}$. In addition, gas in a galaxy of mass
$M>M_{\rm min}$ undergoes an additional starburst if the galaxy merges
with a second galaxy whose mass is larger than $M/2$. This second
star-burst is assumed to be always Pop-II.

We may use our calculation of the evolution of $d\rho_{\rm enrich}/dM$
to calculate the Pop-II SF rate in neutral regions.  
\begin{eqnarray}
\nonumber 
\frac{d\Gamma_{\rm enrich}}{dt}(z)&=&\int_{M_{\rm
min}/2}^{\infty} dM\int_{\max(0,M_{\rm min}-M)}^{\min(M,M_{\rm
min})}d\Delta M \left.\frac{d^2N_{\rm mrg}}{d\Delta
Mdt}\right|_{M}\frac{dt}{dz}\frac{dn_{\rm ps}}{dM}\\ 
\nonumber
&&\hspace{10mm}\times\left(\left[\Theta(M_{\rm
min}-M)\epsilon\eta\frac{\frac{d\rho_{\rm enrich}}{dM}}{\frac{dn_{\rm
ps}}{dM}}\right] + \left[\epsilon\eta\frac{\frac{d\rho_{\rm
enrich}}{d\Delta M}}{\frac{dn_{\rm ps}}{d\Delta M}}\right]\right)\\
\nonumber &&+\int_{M_{\rm min}}^{\infty} dM\int_{M/2}^{M}d\Delta M
\left.\frac{d^2N_{\rm mrg}}{d\Delta
Mdt}\right|_{M}\frac{dt}{dz}\frac{dn_{\rm ps}}{dM}\\
&&\hspace{10mm}\times\epsilon\eta(1-\epsilon\eta)\left[\Theta(M-M_{\rm
min})M + \Theta(\Delta M-M_{\rm min})\Delta M\right] 
\end{eqnarray}
The first term is for SF where gas crossing $M_{\rm min}$ undergoes a
starburst. The second term corresponds to the secondary starbursts (in
gas that has already undergone a burst of SF) that follow major
mergers. Note that while the first term does not contribute in regions
of the IGM that are not yet enriched, this second term can contribute
at all regions of IGM that have not been reionized.

Finally, we calculate the Pop-II SF rate in regions of the IGM that
have been reionized.  
\begin{eqnarray} 
\nonumber 
\frac{d\Gamma_{\rm
enrich,ion}}{dt}(z)&=& \int_{M_{\rm reion}/2}^{\infty}
dM\int_{\max(0,M_{\rm reion}-M)}^{\min(M,M_{\rm reion})}d\Delta M
\left.\frac{d^2N_{\rm mrg}}{d\Delta
Mdt}\right|_{M}\frac{dt}{dz}\frac{dn_{\rm ps}}{dM}\\ \nonumber
&&\hspace{10mm}\times\left(\left[\Theta(M_{\rm
reion}-M)\epsilon\eta\left[f_{\rm b}M\right]\right] +
\epsilon\eta\left[f_{\rm b}\Delta M\right]\right)\\ \nonumber
&&+\int_{M_{\rm min}}^{\infty} dM\int_{M/2}^{M}d\Delta M
\left.\frac{d^2N_{\rm mrg}}{d\Delta
Mdt}\right|_{M}\frac{dt}{dz}\frac{dn_{\rm ps}}{dM}\\
&&\hspace{10mm}\times\epsilon\eta(1-\epsilon\eta)\left[\Theta(M-M_{\rm
min})f_{\rm b}M+\Theta(\Delta M-M_{\rm min})f_{\rm b}\Delta M\right].
\end{eqnarray} 
Here the 1st term is for SF where gas in the halo undergoes a
starburst when the halo crosses $M_{\rm reion}$. The 2nd term
corresponds to the additional starbursts (in gas within halos above
$M_{\rm min}$ that has already undergone a burst of SF) that follow
major mergers.

The Pop-II SF rates in a neutral (dark-dashed lines) and an ionized
(grey-dashed lines) IGM are shown in the lower right panels of
figures~\ref{fig2} and \ref{fig3} assuming $z_{\rm tran}=17$ with
$\Delta z_{\rm tran}=1.25$, and $z_{\rm tran}=22$ with $\Delta z_{\rm
tran}=2.5$ respectively. As noted by Barkana \& Loeb~(2000), the SF
rate in an ionized universe implied by a SF efficiency of $\eta=0.1$
agrees with observational estimates of $\sim0.1M_\odot/$Mpc$^3/$yr at
$z\sim5$.  The integrated Pop-II SF density follows from
\begin{equation} \Gamma_{\rm enrich}(>z) = \int_\infty^z dz
\frac{dt}{dz}\frac{d\Gamma_{\rm enrich}}{dt}(z).  \end{equation}

\section{reionization of the IGM and the SF history}

In the previous section we computed the SF rate as a function of time
in neutral and ionized regions of the universe for both Pop-II and
Pop-III stars. However, the average SF history of the universe cannot
be computed in isolation from its reionization history, which
specifies the fraction of SF that is in neutral and reionized regions
as a function of redshift. Similarly, the computation of a
reionization history requires specification of a SF history. As a
result the SF and reionization histories must be computed in a
co-dependent way.

\subsection{Modeling the Reionization of Hydrogen}

The simplest estimate of the epoch of reionization is based on the
following considerations.  Given a co-moving density of ionizing
photons $n_\gamma$ in a homogeneous but clumpy medium of comoving
density $n_0$ (where the size of the HII region is much larger than
the scale length of clumpiness), the evolution of the volume filling
factor $Q$ of ionized regions is (Shapiro \& Giroux 1987; Haiman \&
Loeb~1997; Madau et al.~1999; Barkana \& Loeb~2001) 
\begin{equation}
\label{fillfactor} \frac{dQ}{dz} =
\frac{1}{n_0}\frac{dn_\gamma}{dz}-\alpha_{\rm B}\frac{C}{a^3}Qn_{\rm
e}\frac{dt}{dz}, 
\end{equation} 
where $\alpha_{\rm B}$ is the case B recombination coefficient,
$a=1/(1+z)$ is the scale factor, $n_{\rm e}$ is the comoving electron
density, and $C\equiv {\langle n_0^2\rangle}/{\langle n_0\rangle^2}$
is the clumping factor.  This equation describes statistically the
transition from a fully neutral universe to a fully ionized one, and
yields reionization redshifts for hydrogen of around 7-12 for fiducial
parameters. However, large uncertainties arise in both the source term
and in the value of the clumping factor (because more rapid
recombinations lead to a slower evolution of $Q$).

A more realistic description of reionization in a clumpy medium is
provided by the model of Miralda-Escude et al.~(2000). In what
follows, we draw primarily from their prescription and refer the
reader to the original paper for a detailed discussion of its
motivations and assumptions. The model assumes that reionization
progresses rapidly through islands of lower density prior to the
overlap of individual cosmological ionized regions. Following overlap,
the remaining regions of high density are gradually ionized. It is
therefore hypothesized that at any time, regions with gas below some
critical overdensity $\Delta_{\rm i}\equiv
{\rho_{i}}/{\langle\rho\rangle}$ are ionized while regions of higher
density are not.  The mass fraction $F_{\rm M}(\Delta_{\rm i})$ (or
equivalently $\Delta_{\rm i}$) therefore evolves according to the
equation 
\begin{equation} 
\label{postoverlap} 
\frac{dF_{\rm
M}(\Delta_{\rm i})}{dz} =
\frac{1}{n_0}\frac{dn_\gamma}{dz}-\alpha_{\rm B}\frac{R(\Delta_{\rm
i})}{a^3}n_{\rm e}\frac{dt}{dz}.  
\end{equation} 
This equation assumes that all ionizing photons are absorbed shortly
after being emitted, so that there is no background ionizing field,
and no loss of ionizing photons due to redshift. We therefore
implicitly assume that the mean free path of ionizing photons is much
smaller than the Hubble length.  This should be valid at redshifts not
too much smaller than the overlap redshift.

The integration of equation~(\ref{postoverlap}) requires knowledge of
$P_{\rm V}(\Delta)$.  Miralda-Escude et al.~(2000) found that a good
fit to the volume weighted probability distribution for the density as
seen in N-body simulations has the functional form 
\begin{equation}
P_{\rm
V}(\Delta)d\Delta=A\exp{\left[-\frac{(\Delta^{-2/3}-C_0)^2}{2(2\delta_0/3)^2}
\right]}\Delta^{-\beta}d\Delta, 
\end{equation} 
with $\delta_0=7.61/(1+z)$ and $\beta=2.23$, 2.35 and 2.48, and
$C_0=0.558$, 0.599 and 0.611 at $z=2$, 3 and 4. At $z=6$ they assume
$\beta=2.5$, which corresponds to the distribution of densities of an
isothermal sphere, and solve for $A$ and $C_o$ by requiring the mass
and volume to be normalized to unity. We repeat this procedure to find
$P_{\rm V}(\Delta)$ at higher redshifts. The proportionality of
$\delta_0$ to the scale factor is expected for the growth of structure
in an $\Omega_{\rm m}=1$ universe or at high redshift otherwise, and
its amplitude should depend on the amplitude of the power-spectrum.
The simulations on which the distribution in Miralda-Escude et
al.~(2000) was based assumed $\Omega_{m}=0.4$ in matter,
$\Omega_\Lambda=0.6$ in a cosmological constant and $\sigma_8=0.79$,
close to the values used in this paper.

Equation~(\ref{postoverlap}) provides a good description of the
evolution of the ionization fraction following the overlap of
individual ionized bubbles, because the ionization fronts are exposed
to the mean ionizing radiation field. However, prior to overlap, the
prescription is inadequate, due to the large fluctuations in the
intensity of the ionizing radiation. A more accurate model to describe
the evolution of the ionized volume prior to overlap was suggested by
Miralda-Escude et al.~(2000). In our notation the appropriate equation
is 
\begin{equation} 
\frac{d[QF_{\rm M}(\Delta_{\rm c})]}{dz} =
\frac{1}{n^0}\frac{dn_{\gamma}}{dz} - \alpha_{\rm
B}(1+z)^3R(\Delta_{\rm c})n_{\rm e}Q\frac{dt}{dz}.  
\end{equation}
or 
\begin{equation} 
\label{preoverlap} 
\frac{dQ}{dz} = \frac{1}{n^0
F_{\rm M}(\Delta_{\rm c})}\frac{dn_{\gamma}}{dz} -
\left[\alpha_{\rm B}(1+z)^3R(\Delta_{\rm c})n_{\rm e}\frac{dt}{dz}
+ \frac{dF_{\rm M}(\Delta_{\rm c})}{dz}\right]\frac{Q}{F_{\rm
M}(\Delta_{\rm c})}.  
\end{equation} 
In this expression, $Q$ is redefined to be the volume filling factor
within which all matter at densities below $\Delta_{\rm c}$ has
been ionized. Within this formalism, the epoch of overlap is precisely
defined as the time when $Q$ reaches unity. However, we have only a
single equation to describe the evolution of two independent
quantities $Q$ and $F_{\rm M}$. The relative growth of these depends
on the luminosity function and spatial distribution of the
sources. The appropriate value of $\Delta_{\rm c}$ is set by the
mean free path of the ionizing photons. More numerous sources can
attain overlap for smaller values of $\Delta_{\rm c}$. Assuming
$\Delta_{\rm c}$ to be constant with redshift, we find that results
do not vary much (less than 10\% in the optical depth to electron
scattering) for values of $\Delta_{\rm c}$ ranging from a few to a
few tens. At high redshift, these $\Delta_{\rm c}$ correspond to
mean free paths comparable to the typical separations between galaxies
or quasars. Unless otherwise specified we assume $\Delta_{\rm
crit}=10$ (which lies between the values for galaxies and quasars)
throughout the remainder of this paper.

\subsection{Effects of Minihalos On Reionization}

The presence of minihalos effects the progression of reionization, by
delaying the buildup of the ionizing radiation field within ionized
regions.  Many of the ionizing photons will be used to photo-evaporate
the minihalos rather than to progress reionization within
IGM. Moreover, the finite time over which this evaporation occurs
allows SF to proceed in halos crossing $M_{\rm min}$ for a short time
following the reionization of a region of IGM. In the presence of an
intense ionizing background radiation field, a large fraction of the
gas in halos with virial temperatures below $T_{\rm vir}\sim10^4$K is
photo-evaporated (Barkana \& Loeb~2002; Iliev, Shapiro, \& Raga, 2005;
Ciardi et al., 2005) on the halo dynamical time $(\sim0.1H^{-1})$. Two
additions should therefore be made to the above description arising
from the effect of minihalos.  First, we need to take into account
Pop-III SF within ionized regions due to halos crossing the H-cooling
threshold which have not yet been photo-evaporated.  A simple way to
approximate the effect of this additional SF is to reduce the cold gas
in minihalos, and therefore the resulting SF rate corresponding to
crossing the H-cooling threshold by a factor of
$0.5(Q(t)-Q(t-dt))/Q(t)$ at time $t$. Here $dt\sim0.1H^{-1}$ and the
pre-factor of 0.5 is approximately the average remaining neutral gas
fraction in all minihalos which become exposed to ionizing photons
between $dt$ and zero years prior to a redshift $z$, while the ratio
term is just the fraction of minihalos that become exposed to ionizing
radiation within the preceding $dt$ years.

Second, while the above consideration enhances the SF in ionized IGM,
the minihalos will delay further reionization for a period lasting for
the photo-evaporation timescale through screening of the ionizing
sources by minihalos (Barkana \& Loeb~2002). We therefore need to
estimate the fraction of photons that are absorbed by minihalos rather
than contributing to the reionization of the IGM. This estimate may be
made as follows. An ionizing photon will travel a distance
$\lambda_{\rm gas}$ through ionized regions of IGM before encountering
neutral IGM. On the other hand, an ionizing photon will travel a
different distance $\lambda_{\rm halo}$ before encountering a minihalo
that contains cold gas. The fraction of ionizing photons produced in
ionized regions that reionize neutral IGM rather than intercepting
minihalos prior to their evaporation is therefore approximated by
\begin{equation} f_{\rm mini}=\exp\{-\lambda_{\rm gas}/\lambda_{\rm
halo}\}.  \end{equation}

\begin{table*}[t]
\begin{center}
\begin{small}
\caption{\label{tab1} Parameters for models presented in Figures~\ref{fig4}-\ref{fig9}.}
\begin{tabular}{cccccccccccccc}
\hline 
                  & SF      & $z_{\rm tran}$ & $\Delta z_{\rm tran}$ & $\Delta_{\rm c}$ & $f_{\rm mix}$ & \multicolumn{4}{c}{ $f_{\rm enrich},f_{\rm primord}$} & $\tau_{\rm es}$ \\
\hline 
Figure~\ref{fig4} & Case-A  &   17           &  1.25                 & 20             &   10\%    &   0.007,0.01&0.006,0.03&0.004,0.10&0.008,0.30  &  0.19 0.18 0.15 0.12 \\
Figure~\ref{fig5} & Case-A  &   22           &  2.5                 & 20              &   100\%    &   0.009,0.01&0.009,0.03&0.010,0.10&0.009,0.30 &  0.20 0.18 0.14 0.11\\
Figure~\ref{fig6} & Case-B  &   17           &  1.25                 & 20             &   100\%    &   0.122,0.03&0.093,0.10&0.058,0.30&0.037,1.0 &   0.13 0.09 0.07 0.06\\
Figure~\ref{fig7} & Case-B  &   22           &  2.5                 & 20              &   100\%     &   0.130,0.03&0.115,0.10&0.090,0.30&0.067,1.0 & 0.12 0.08 0.06 0.05 \\
Figure~\ref{fig8} & Case-A  &   17           &  1.25                 & 10             &   10\%      &   0.015,0.01&0.015,0.03&0.015,0.10&0.012,0.30 &  0.21 0.18 0.15 0.12 \\
Figure~\ref{fig9} & Case-A  &   22           &  2.5                 & 10              &   100\%     &   0.018,0.01&0.018,0.03&0.021,0.10&0.022,0.30 &   0.21 0.18 0.14 0.11\\ \hline
\end{tabular}
\end{small}
\end{center}
\end{table*}

%FIGURE 4 
\begin{figure*}[t] 
\epsscale{.8} 
\plotone{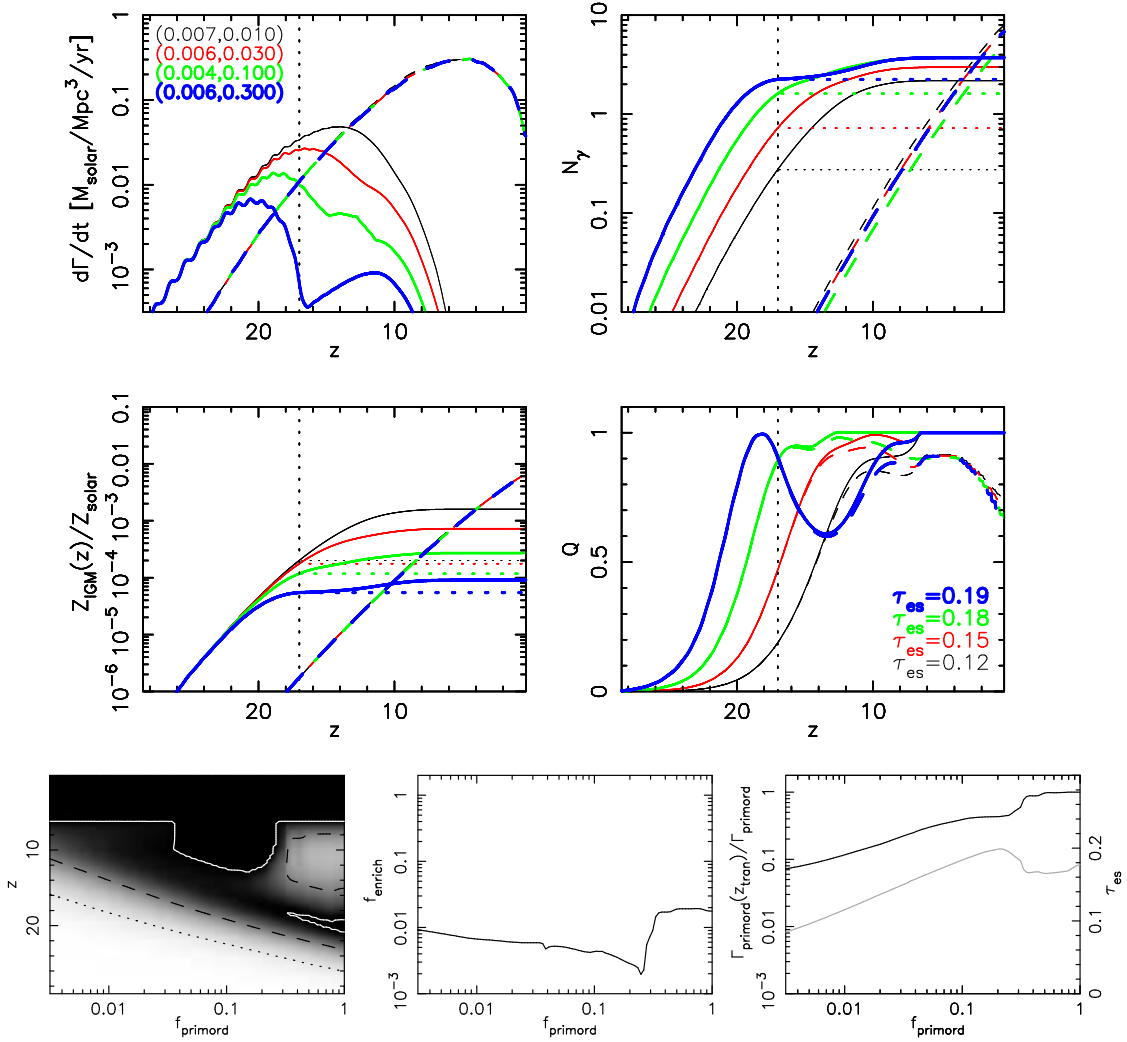}
\caption{\label{fig4} {Reionization and SF histories for the case
where the redshift at which the IGM became metal enriched is taken to be
$z_{\rm tran}=17$ with a width of $\Delta z_{\rm tran}=1.25$ and the
metal mixing efficiency is $f_{\rm mix}=10\%$.  The epoch $z_{\rm tran}$ is marked by
the vertical dotted line. The top-left panel shows the differential SF
rates for Pop-III (solid lines) and Pop-II stars (dashed curves). In
the top-right panel, we plot the cumulative number of ionizing photons
per baryon that have escaped into the IGM by redshift $z$ for Pop-III
(solid curves) and Pop-II (dashed curves) stars. The dotted curve
marks number of ionizing photons produced by Pop-III stars at $z_{\rm
tran}$. The middle left panel shows the contributions to the mean
metallicity of the IGM from Pop-III (solid curves) and Pop-II (dashed
curves) SF. The right hand panel shows the reionization history as the
evolution of the volume averaged (solid line) and mass averaged
ionization fractions (dotted line). The lines of different thickness
show histories with different combinations of $(f_{\rm enrich},f_{\rm
primord})$ respectively.  In each panel the thicker lines correspond
to larger values of $f_{\rm primord}$. The lower three panels
represent results for series of reionization histories. The value of
$f_{\rm enrich}$ that results in a final overlap at $z=6.5$ (or as
near as possible at $z>6.5$) is plotted as a function of $f_{\rm
primord}$ in the lower middle panel. The lower left panel presents
grey-scale and contours showing the level of overlap as a function of
redshift and $f_{\rm primord}$. The grey-scale shows the level of
ionization (with black representing high ionization). The dotted,
dashed and solid lines correspond to $Q=0.2$, 0.5 and .999
respectively. In the lower right panel we plot the optical depth
$\tau_{\rm es}$ as a function of $f_{\rm primord}$ (grey line). The
dark line in this panel shows the ratio of the amount of Pop-III SF
already completed at $z_{\rm tran}$ to the total Pop-III SF.  }}
\end{figure*}

%FIGURE 5
\begin{figure*}[t]
\epsscale{.8}
\plotone{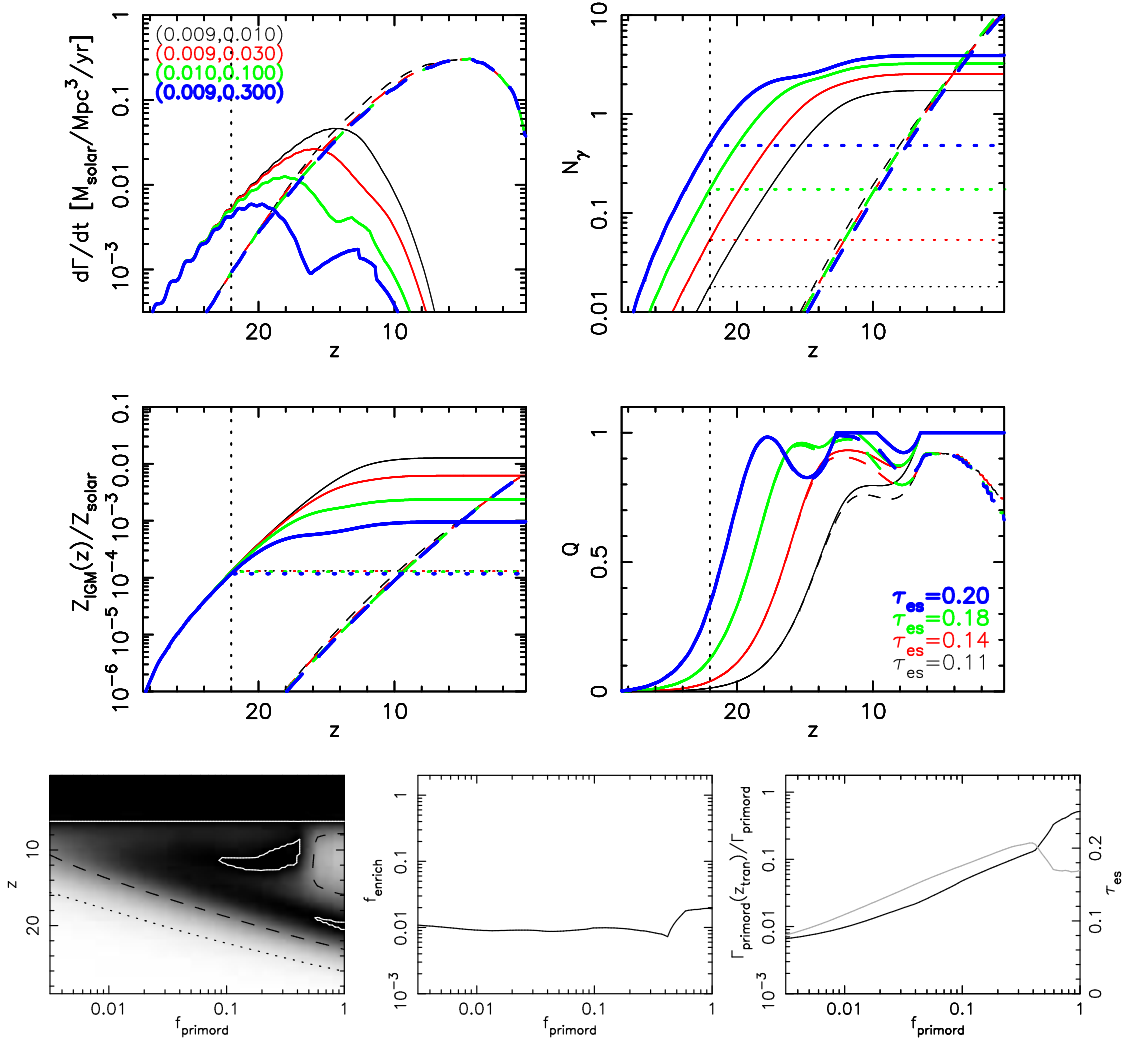}
\caption{\label{fig5} As per Figure~\ref{fig4} but with $z_{\rm tran}=22$ and $\Delta z_{\rm tran}=2.5$, and $f_{\rm mix}=100\%$.}
\end{figure*}

%FIGURE 6
\begin{figure*}[t]
\epsscale{.8}
\plotone{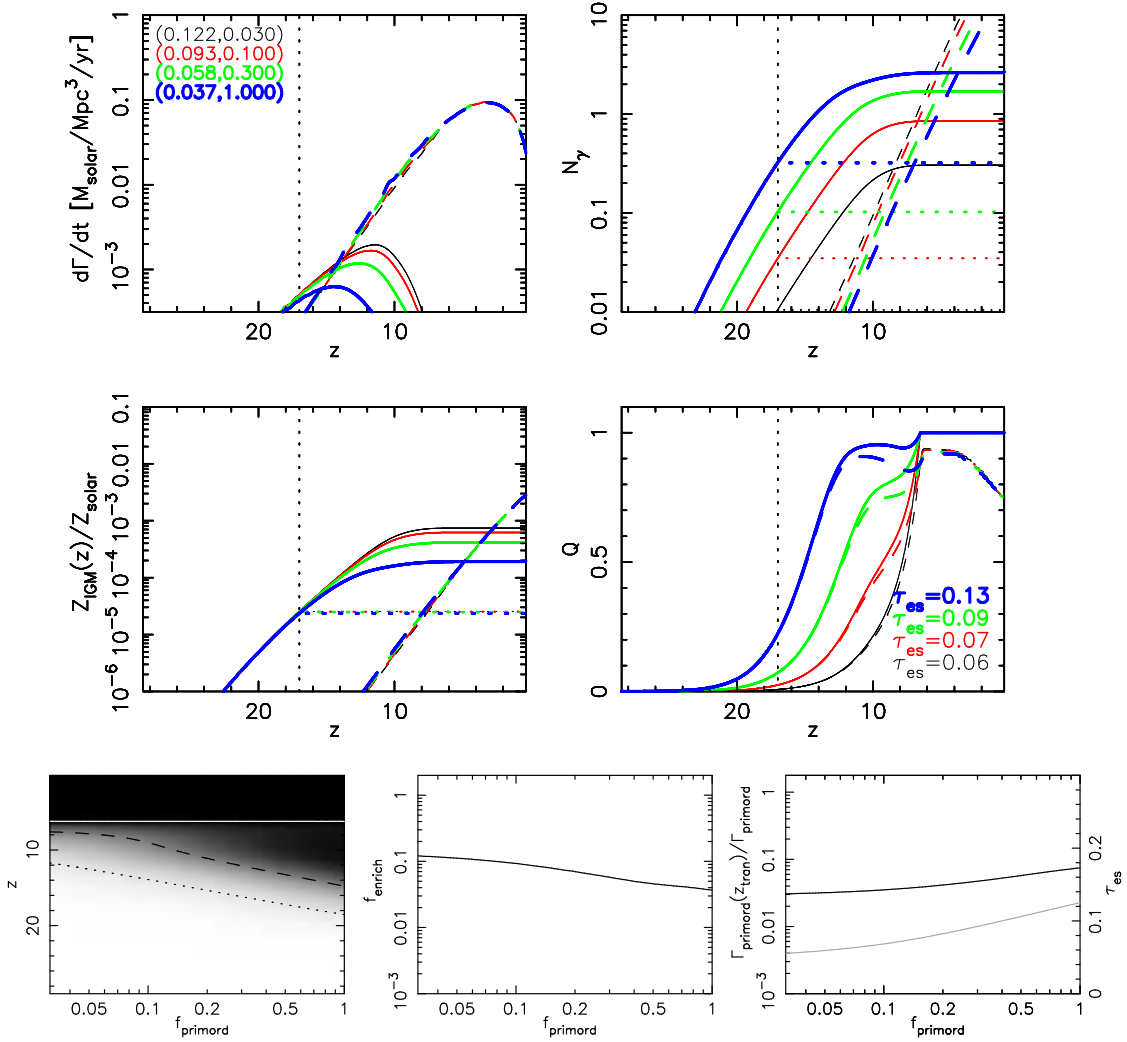}
\caption{\label{fig6} As per Figure~\ref{fig4} but with case-B SF, and $f_{\rm mix}=100\%$.}
\end{figure*}

%FIGURE 7
\begin{figure*}[t]
\epsscale{.8}
\plotone{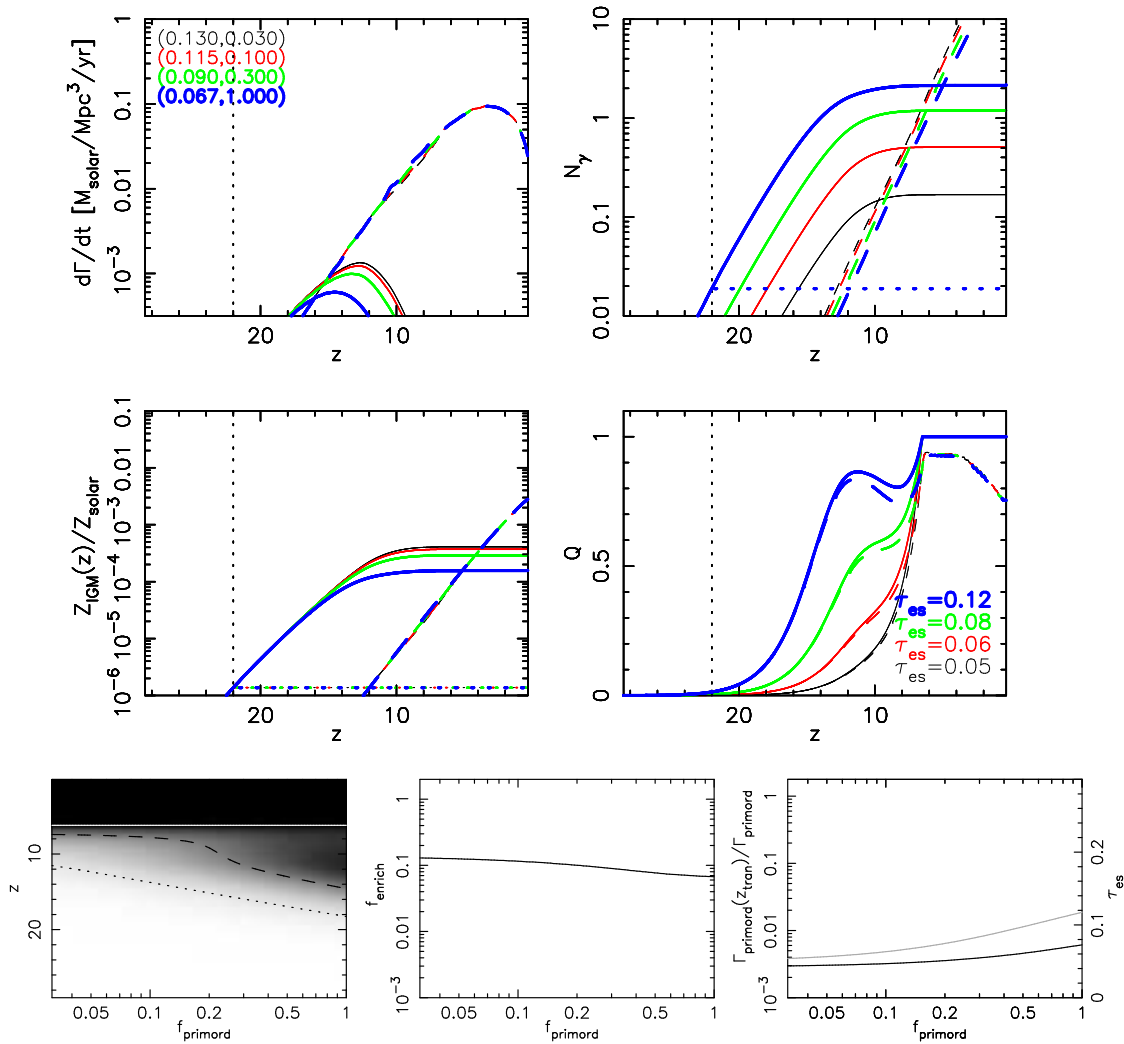}
\caption{\label{fig7} As per Figure~\ref{fig4} but with $z_{\rm tran}=22$ and $\Delta z_{\rm tran}=2.5$, and case-B SF, and $f_{\rm mix}=100\%$.}
\end{figure*}

%FIGURE 8
\begin{figure*}[t]
\epsscale{.8}
\plotone{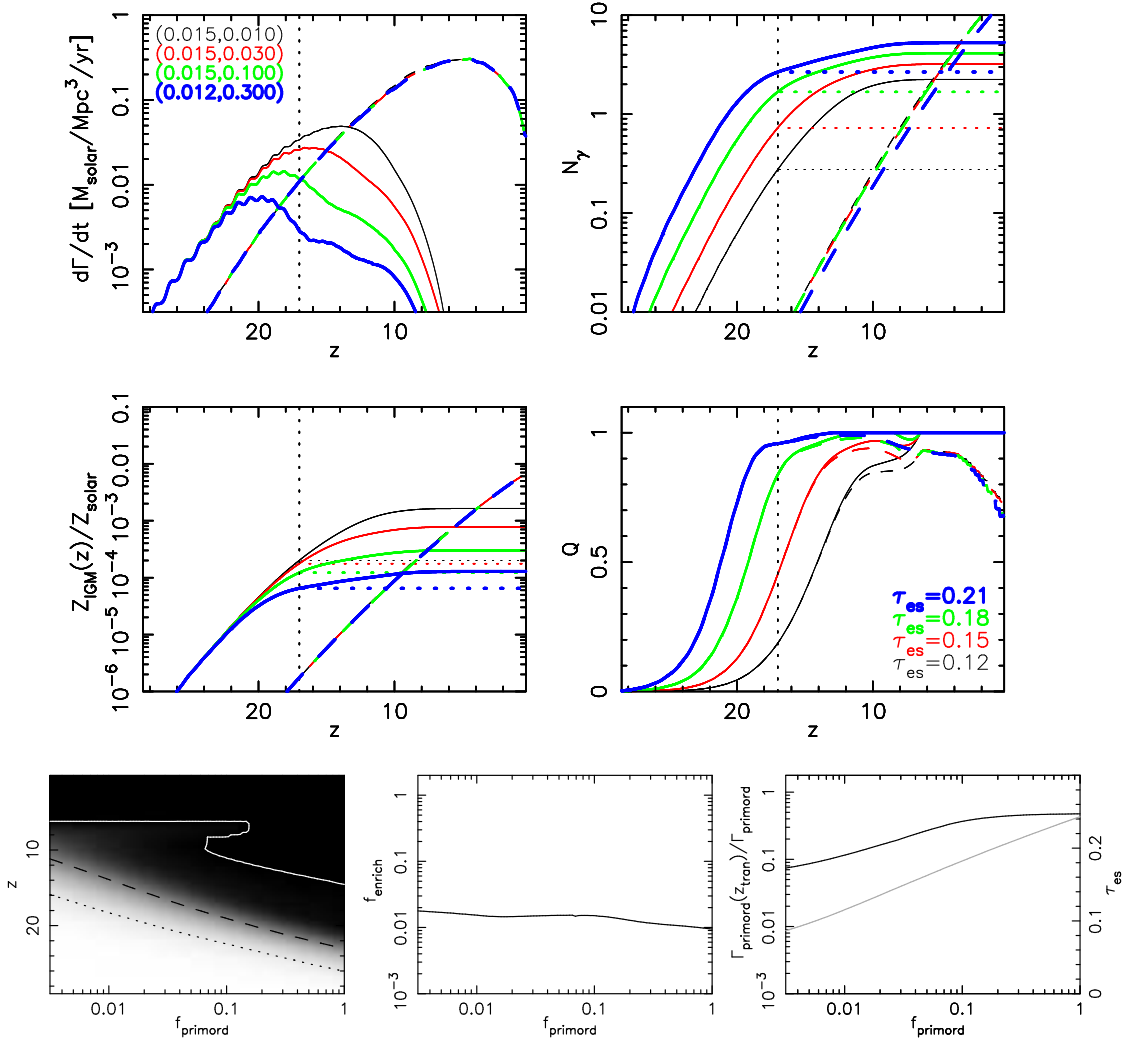}
\caption{\label{fig8} As per Figure~\ref{fig4} but with $\Delta_{\rm c}=20$.}
\end{figure*}

%FIGURE 9
\begin{figure*}[t]
\epsscale{.8}
\plotone{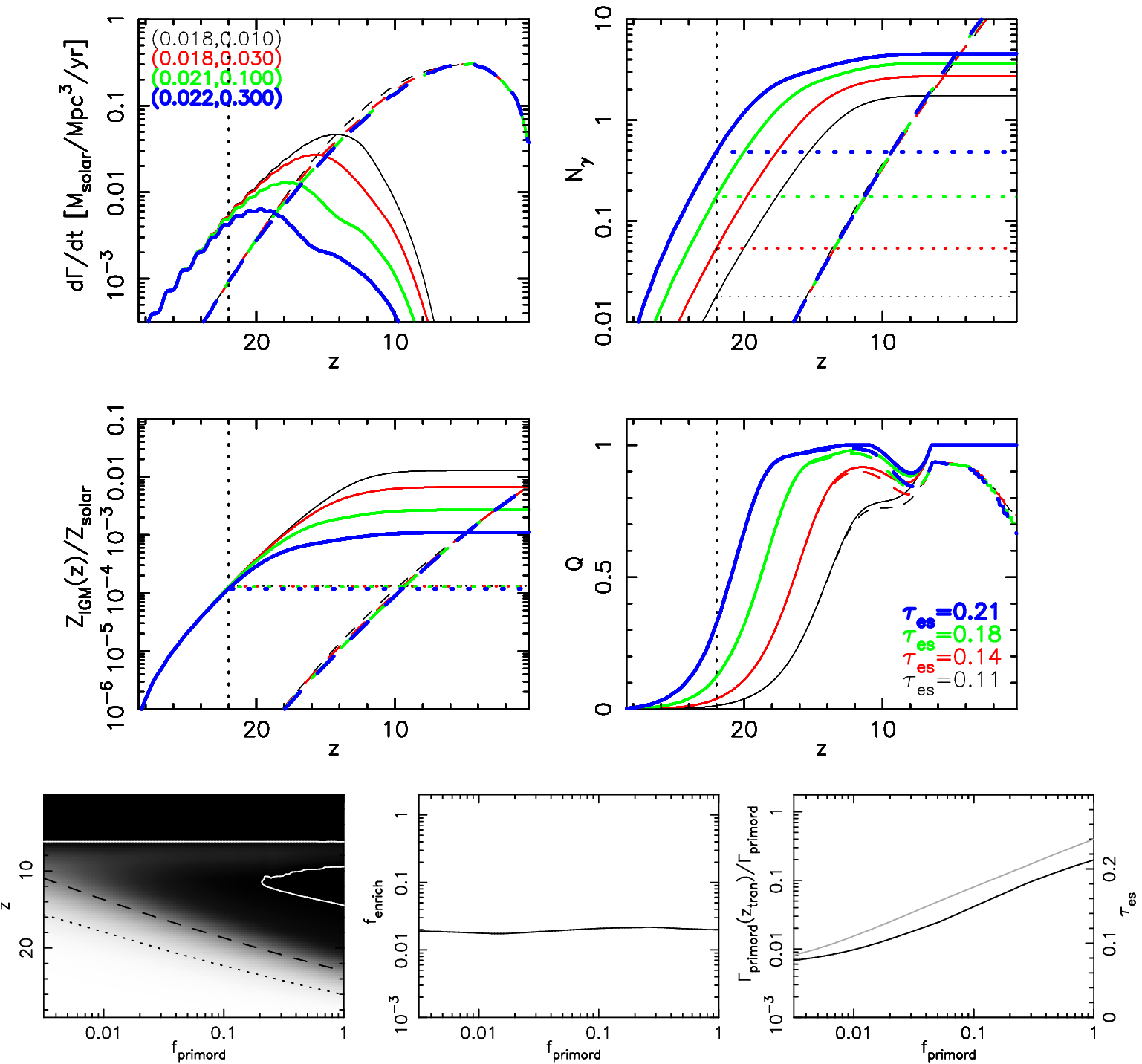}
\caption{\label{fig9} As per Figure~\ref{fig4} but with $z_{\rm tran}=22$ and $\Delta z_{\rm tran}=2.5$, $\Delta_{\rm c}=20$, and $f_{\rm mix}=100\%$.}
\end{figure*}

The length $\lambda_{\rm gas}$ may be easily computed within the
formalism described. An ionizing photon will typically travel a
distance $\lambda_{\rm gas}=60\mbox{km}\,\mbox{s}^{-1}H^{-1}[1-F_{\rm
V}]^{-2/3}$, where $F_{\rm V}$ is the fraction of the IGM at
overdensities below $\Delta_{\rm c}$ (Miralda-Escude et al.~2000).
On the other hand, to compute the cross-section for interception of a
minihalo that contains cold gas we need to find the fraction of gas in
minihalos at a redshift $z$ that was accreted to within progenitor
minihalos within the previous interval $0.1H^{-1}$. This quantity may
be computed from the extended Press-Schechter formalism (Lacey \&
Cole~1993). For a halo of mass $M$ at redshift $z$, the fraction of
the halo mass which by some higher redshift $z_2$ had already
accumulated in halos is 
\begin{equation} 
F_{\rm g}(z,z_2) =
\mbox{erfc}\left[\frac{1.69/D(z_2)-1.69/D(z)}{\sqrt{2[S(M_{\rm
J})-S(M)]}}\right], 
\end{equation} 
where $D(z)$ is the linear growth factor at redshift $z$, $S(M)$ is
the variance on mass scale $M$ (defined using the linearly
extrapolated power spectrum at $z=0$), and $M_{\rm J}(z_2)$ is the
Jeans mass at $z_2$. Given the virial radius for a halo $R_{\rm
vir}(M)$ of mass $M$ (Barkana \& Loeb~2001) and the
Press-Schechter~(1974) mass function ($dn/dM$) for the number density
of halos we may then find the length $\lambda_{\rm halo}$
\begin{equation} 
\lambda_{\rm halo} = \left[\int_{M_{\rm J}}^{M_{\rm
min}} dM \pi R_{\rm vir}^2\frac{dn}{dM} [1-F_{\rm g}(z,z+\delta
z)]\right]^{-1}, 
\end{equation} 
where $\delta z=0.1H^{-1}(dt/dz)^{-1}$.

\subsection{Co-Evolution of Reionization, Star Formation and Metal Enrichment}

We next describe the co-evolution of the reionization and SF
histories. Having specified a transition redshift, we computed in the
previous section the Pop-III ($\frac{d\Gamma_{\rm primord}}{dt}$) and
Pop-II ($\frac{d\Gamma_{\rm enrich}}{dt}$) SF rates in neutral
regions, as well as the Pop-II SF rates in ionized regions
($\frac{d\Gamma_{\rm enrich,ion}}{dt}$). The ionizing photon
production rate then follows from 
\begin{eqnarray} 
\label{history}
\frac{dn_\gamma}{dz} &=& \frac{dn_{\gamma,{\rm quasars}}}{dz} +
\frac{dt}{dz}\frac{M_\odot}{m_{\rm p}}\left[(1-Q_{\rm
heat})\left(N_{\rm enrich}f_{\rm enrich}\frac{d\Gamma_{\rm
enrich}}{dt}(z) + N_{\rm primord}f_{\rm esc}\frac{d\Gamma_{\rm
primord}}{dt}(z)\right) \right.\\ 
\nonumber &&\hspace{-15mm}
+\left.Q_{\rm heat} f_{\rm mini} \left(f_{\rm enrich}N_{\rm
enrich}\frac{d\Gamma_{\rm enrich,ion}}{dt}(z)+ \frac{1}{2}\frac{Q_{\rm
heat}-Q_{\rm heat}(z+\delta z)}{Q_{\rm heat}}N_{\rm primord}f_{\rm
esc}\frac{d\Gamma_{\rm primord}}{dt}(z)\right)\right], 
\end{eqnarray}
where $m_{\rm p}$ is the proton mass, $N_{\rm primord}$ and $N_{\rm
enrich}$ are the ionizations per baryon, and $f_{\rm primord}$ and
$f_{\rm enrich}$ are the escape fractions for Pop-III and Pop-II stars
respectively. The quantity $Q_{\rm heat}$ is the maximal value that
$Q$ has attained at any time in the reionization history.  A region
that has been reionized may recombine in the absence of an ionizing
radiation field, but the timescale for cooling is longer than the
recombination timescale for $\Delta_i\ge 10$.  The inclusion of
$Q_{\rm heat}$ ensures that the SF rate in regions that have
recombined remains the same as for an ionized region should that
region return to a neutral state. The term $\frac{dn_{\gamma,{\rm
quasars}}}{dz}$ accounts for the contribution of quasars to the
ionizing flux. Quasars are included in the model as described in
Wyithe \& Loeb~(2003a), with an updated quasar luminosity function
model (Wyithe \& Loeb~2003b). If treated as the sole ionizing sources,
we find that quasars reionize cosmic hydrogen at around
$z\sim5$. Wyithe \& Loeb~(2003c) found that helium became reionized by
quasars at $z\sim3.5$, in agreement with observation (e.g. Theuns et
al.~2002).  The coupled evolution of equations~(\ref{postoverlap}),
(\ref{preoverlap}) and (\ref{history}) yield the SF and reionization
histories.

We may also estimate the mean level of metal enrichment in the
IGM. This quantity is important because massive Pop-III stars are
expected to form only out of gas that is not enriched to a metallicity
above $Z_{\rm crit}\sim 10^{-4}$Z$_\odot$ (Bromm et al. 2001a).  We
estimate the mean metallicity at a redshift $z$ as 
\begin{equation}
Z_{\rm IGM}(z)=f_{\rm mix}\frac{Y}{\omega_{\rm SN}}\int_{z}^\infty dz'
\frac{dF_\star}{dz'}, 
\end{equation} 
where $\omega_{\rm SN}$ is the stellar mass that must be produced per
supernova, $Y$ is the mass of metals produced per supernova, and
$f_{\rm mix}$ is the fraction of metals that escape from galaxies into
the IGM in supernova-driven outflows (e.g. Madau, Ferrara \& Rees
2001; Thacker, Scannapieco \& Davis 2002). The value of $f_{\rm mix}$
may be large because the early dwarf galaxies had shallow potential
wells. For a massive Pop-III IMF, Furlanetto \& Loeb~(2003) estimate
$\omega_{\rm SN}\sim 462M_\odot$ and $Y\sim70.5M_\odot$. We note that
this number is very sensitive to the stellar mass function, because
stars with masses $140\la M\la 260M_\odot$ are expected to eject their
metals while more massive stars collapse to form black holes (Heger \&
Woosley~2002).

Different reionization histories lead to different column depths of
free electrons along a line-of-sight to recombination. An important
probe of the reionization history is therefore provided by the
observation by the WMAP satellite of a large optical depth to electron
scattering of CMB photons. The optical depth to electron scattering
$\tau_{\rm es}$, depends on the mass filling factor ($QF_{\rm M}$) and
is evaluated according to 
\begin{equation} 
\tau_{\rm es} = \int_0^{1000}dz\frac{cdt}{dz}\sigma_{T}QF_{\rm M}n_{\rm H}^{\rm
0}(1+z)^3, 
\end{equation} 
where $\sigma_{T}=6.652\times 10^{-25}~{\rm cm^2}$ is the Thomson
cross-section.

\subsection{Results: The Reionization History of Hydrogen  }

Figures~\ref{fig4}-\ref{fig9} show several examples of the
reionization history, which explore a wide range of the various inputs
to the model.  These inputs are summarized in Table~\ref{tab1}.  In
each case we present 4 histories in the upper set of panels. The input
values of $f_{\rm primord}$, together with the value of $f_{\rm
enrich}$ that leads to a final reionization at $z=6.5$ are also listed
in table~\ref{tab1}. The thicker lines correspond to histories with
larger values of $f_{\rm primord}$. Note here that the escape fraction
is degenerate with the SF efficiency (for $\eta\ll1$).  We have chosen
$\eta=10\%$ in presenting our SF histories, however the values of
escape fraction can be adjusted in proportion for different choices of
$\eta$.

The total SF rates in Pop-II (solid lines) and Pop-III (dashed lines)
stars are plotted in the upper left panels of
Figures~\ref{fig4}-\ref{fig9}. Feedback from reionization quenches
SF. However, our histories show that the additional Pop-III SF remains
significant at $z<z_{\rm tran}$ in all cases, and peaks at $z<z_{\rm
tran}$ in many cases.  Pop-II stars begin to dominate the global SF
rate only around the time of the peak in the Pop-III SF rate. We also
find that Pop-III SF at redshifts below $z_{\rm tran}$ forms the
dominant contribution to the total integrated Pop-III SF rate in all
models. This may be seen through plots of the integrated ionizing
photon production as a function of redshift (upper right panels of
Figures~\ref{fig4}-\ref{fig9}). The horizontal dotted lines show the
level of ionizing photon production at $z=z_{\rm tran}$. The fraction
of ionizing photon production by Pop-III at $z<z_{\rm tran}$ is larger
than 90\% in nearly all cases, and can be as large as
99.9\%. Furthermore, Pop-III SF can dominate the total ionizing photon
production at $z>6.5$ for models with large $\tau_{\rm es}$. We note
that the comparison of SF in an ionized and in a neutral IGM (see
Figures~\ref{fig2}-\ref{fig3}) suggests that reionization of the IGM
should be accompanied by a dip in the SF rate as seen in the
calculations of Barkana \& Loeb~(2000). However, the SF histories that
are computed in concert with a reionization history do not show this
dip. The reason is that an extended reionization history washes out
this feature on average, though it may still be observed in isolated
regions.

For each of the reionization histories in
Figures~\ref{fig4}-\ref{fig9} we plot $Z_{\rm IGM}(z_{\rm
tran})/Z_\odot$ (middle left panels).  We have chosen $f_{\rm mix}$
(see values in Table~\ref{tab1}) so as to obtain
$Z/Z_\odot\sim10^{-4}$; close to the threshold deduced by Bromm et al.
(2001a) at $z_{\rm tran}$. Note that $Z/Z_\odot$ is degenerate with
$\eta$ in addition to $f_{\rm mix}$. However, this simple estimate
demonstrates that metal enrichment of the IGM by Pop-III stars is
consistent with our assumed $z_{\rm tran}$ (see also Mackey, Bromm \&
Hernquist~2002). The possible exception is for Case-B SF with $z_{\rm
tran}=22$. In this example the integrated SF rate by $z_{\rm tran}$ is
not sufficient to enrich the IGM. Finally, the parameter $\Delta_{\rm
c}$ is unspecified a-priori. We have chosen $\Delta_{\rm c}=10$, with
the exception of Figures~\ref{fig8} and \ref{fig9}. Comparison of
Figures~\ref{fig8} and \ref{fig9} with Figures~\ref{fig4} and
\ref{fig5} demonstrate that while the details of the reionization
history depend on the value assumed for $\Delta_{\rm c}$, the
qualitative nature of the predicted histories and the values of
$\tau_{\rm es}$ are similar. We therefore find that our conclusions
are not sensitive to the choice of $\Delta_{\rm c}$.

We find that either Pop-III stars substantially reionize the universe,
or that the fraction of Pop-III SF which takes place following the
transition redshift is large. Either way the conclusion is that
Pop-III stars played a major role in the reionization history of the
universe. In particular if the universe did not become substantially
reionized prior to $z_{\rm tran}$, then the large majority of Pop-III
SF takes place at redshifts below $z_{\rm tran}$, in contrast to
previous modeling. Each of the histories shown in
Figures~\ref{fig4}-\ref{fig9} was chosen so that overlap occurs at
$z\sim6.5$ as suggested by the spectra of high redshift SDSS
quasars. However, different reionization histories result in different
densities of electrons as a function of redshift.  The values of
$\tau_{\rm es}$ for the various models are listed in the middle right
panels of Figures~\ref{fig4}-\ref{fig9} and in Table~\ref{tab1}.  WMAP
suggested $\tau_{\rm es}>0.1$. We see that only a small value of $\eta
f_{\rm primord}\ga10^{-3}$ for Case-A, or $\eta f_{\rm
primord}\ga10^{-2}$ for Case-B is required to obtain this large
optical depth, even where the enrichment of the IGM occurred at a
redshift as high as $z_{\rm tran}=22$. Indeed the conclusion that
Pop-III stars contributed significantly to reionization is quite
insensitive to the transition redshift for $\eta f_{\rm
primord}\ga10^{-3}$, so long as Pop-III stars had a top-heavy IMF.

The lower three panels of Figures~\ref{fig4}-\ref{fig9} represent
results for a series of reionization histories. For each value of the
primordial escape fraction ($f_{\rm primord}$), we have found the
value of $f_{\rm enrich}$ that results in a final overlap at $z=6.5$
(or as near as possible at $z>6.5$), and which maintains the
reionization until the present day.  In the lower left panel we
present grey-scale and contours showing the level of overlap as a
function of redshift and $f_{\rm primord}$. The grey-scale shows the
level of ionization (with black representing high ionization). The
dotted, dashed and solid lines correspond to $Q=0.2$, 0.5 and .999
respectively. The value of $f_{\rm enrich}$ corresponding to these
histories is plotted as a function of $f_{\rm primord}$ in the middle
panel. In the right panel we plot the optical depth $\tau_{\rm es}$ as
a function of $f_{\rm primord}$ (grey line). For Case-A, the optical
depth has a maximum near $\tau_{\rm es}\sim0.2$, but the WMAP result
is consistent with all values of $f_{\rm primord}$. For Case-B, the
optical depth has a maximum near $\tau_{\rm es}\sim0.1$. Only larger
values of $f_{\rm primord}$ are consistent with the WMAP result for
Case-B. The dark line in the lower-right panel is the ratio of the
amount of Pop-III SF already completed at $z_{\rm tran}$ to the total
Pop-III SF.  This ratio is small ($\sim0.1$) for small values of
$\tau_{\rm es}$. In Case-A, we find that the ratio can be as small as
0.01 where $z_{\rm tran}=22$ and 0.1 where $z_{\rm tran}=17$. The
ratio is only of order unity if $f_{\rm primord}$ is also of order
unity. In Case-B, we find that the ratio can be even smaller, with
values a factor of 3 lower. The ratio does note become of order unity
for any value of $f_{\rm primord}$ in Case-B.

Interestingly in Case-A a range of histories show two epochs of
reionization. However, contrary to earlier works, the tail end of the
first reionization has not been set by an abrupt transition redshift
$z_{\rm tran}$. Rather the end of the first reionization can come
either before, or after $z_{\rm tran}$ depending on the sum of the
various astrophysics included in the model. A particularly interesting
history is seen where Case-A SF is combined with a large escape
fraction for Pop-III ionizing photons and a large transition redshift
(see Figure~\ref{fig5}). In this example the reionization history
shows three peaks of reionization.  Thus, the claim made earlier by
Furlanetto \& Loeb (2005) that a temporally smooth metallicity
enrichment of the IGM gives rise to only monotonic reionization
process is not borne out in our more detailed calculations in general.

Clearly, the information embedded in the reionization history is very rich. 
It is useful to extract some essential properties with regard
to the CMB experiments such as WMAP, since the three-year WMAP results
are expected to be made public shortly.
The thin black curves in Figures~\ref{fig6} and \ref{fig7} show histories with 
$\tau_{\rm es}=0.06$ and $0.05$ respectively. These histories follow from 
a very small Pop-III contribution to reionization,
suggesting that without Pop-III contribution we expect $\tau_{\rm es}\le 0.05-0.06$.
At the other extreme, when we maximize the contribution from Pop-III
stars to reionization by using a very high (perhaps implausible)
escape fraction (relative to that of Pop-II stars),
a maximum value of $\tau_{\rm es}=0.21$ is possible. This 
suggests that the optical depth reported by the three-year WMAP results should not exceed this value
(which is $1\sigma$ above the central value of the first year WMAP result),
otherwise our understanding of structure formation 
at high redshift would require a dramatic revision.
In a more realistic picture where 
the product of star formation efficiency and escape fraction 
for Pop-III stars is comparable to that for Pop-II stars,
we find that the range $\tau_{\rm es}=0.09-0.12$ 
is most likely, consistent with our previous calculations (Cen 2003a,b).

The three different outcomes described above imply qualitatively different histories. 
In the small $\tau_{\rm es}\le 0.05-0.06$ case without significant Pop-III contribution,
reionization is expected to be rapid with the neutral fraction
quickly rising above $50\%$ at $z\sim 8$.
In the large $\tau_{\rm es}=0.012-0.21$ case with a large Pop-III contribution,
the reionization process could be complex and frequently displays multiple peaks.
Under seemingly more reasonable assumptions regarding the efficiency of
Pop-III stars, there appears to be a reionization plateau 
at $z=7-12$ where the mean neutral fraction holds
in a relatively narrow range of $10\%-30\%$.
Independent constraints of the reionization history are highly desirable.  
Surveys of Ly$\alpha$ emitters (e.g., Malhotra \& Rhoads 2005) and measurements of QSO
Stromgren sphere sizes (e.g., Mesinger, Haiman, \& Cen 2004; Wyithe,
Loeb \& Carilli~2005; Fan et al~2005) may provide powerful
constraints on the evolution of the ionized fraction close to the end
of the reionization epoch.  
In addition future CMB experiments such as Planck
survey might be able to distinguish (e.g., Holder et al. 2003) between
the various histories and if so, it should yield some critical
information on SF processes in the high redshift universe.

\section{Conclusion}

We show that the collapsed fraction of primordial gas that is trapped
in virialized minihalos at the redshift ($z_{\rm tran}$) when the IGM
becomes enriched with metals is larger than the collapsed fraction of
gas that has already been involved in SF by a factor of a few to a few
tens.  We argue that this virialized gas will not be as easily
enriched by super-galactic winds as the general IGM, because the
strong self-gravity created by dark matter and significant overdensity
with respect to the background gas help stabilize the collapsed gas
against large disruptions by sweeping shockwaves.  Moreover, enriched
gas is expected to have been pre-heated during the process of
enrichment, preventing contamination of virialized primordial gas
during hierarchical growth of mini-halos. Hence, collapsed gas in
virialized minihalos may largely remain metal-free, until those halos
merge to form larger systems of sufficient mass to initiate SF. 

We have followed the evolution of virialized primordial gas through
hierarchical mergers of their host dark matter halos. This enables the
calculation of the Pop-III SF rate arising from the virialized
primordial gas when it crosses the hydrogen cooling threshold.  Our
model for the reionization of the IGM computes the co-evolution of the
reionization and SF histories.  In addition to following the
gas-content of halos below the minimum mass for SF our model addresses
several astrophysical phenomena. These include feedback from
reionization in the form of suppression of galaxy formation in a
photo-ionized IGM, as well as the photo-evaporation of minihalos (both
the screening of ionizing sources and the additional low mass SF in
ionized regions prior to photo-evaporation). We also allow the IGM to
be gradually enriched by metals rather than an abrupt transition.

We find that the Pop-III SF rate peaks substantially later (as low as
redshift $z\sim 10$) in cosmic time than the epoch at which the IGM
became metal enriched, and may extend to $z\sim 6$. Moreover, we find
that the total integrated Pop-III SF is larger, by a factor of a few
to a few tens, than the integrated Pop-III SF at the average redshift
of IGM metal enrichment.  This prolonged epoch of Pop-III SF has a
pronounced effect on the reionization history of cosmic hydrogen.
Despite the varied feedback mechanisms included in our model 
we find that the prolonged
epoch of Pop-III SF leads to complex reionization histories 
which can be multi-peaked under a range of plausible scenarios.  

Observational measurements of the detailed reionization history would
provide critical information on the formation process of first stars.
Specifically, assuming the final overlap redshift to be fixed at
$z=6.5$ (as implied by observations of high redshift quasars), the
contribution of Pop-III stars to reionization can be quantified and
will be tested by three-year WMAP results: (1) if Pop-III stars do not
contribute to reionization, $\tau_{\rm es}\le 0.05-0.06$ and a rapid
reionization at $z\sim 6$ is expected, with the mean neutral fraction
quickly exceeding $50\%$ at $z\sim 8$; (2) if the product of star
formation efficiency and escape fraction for Pop-III stars is
significantly larger than for Pop-II stars, then a maximum
$\tau_{\rm es}=0.21$ is achievable; (3) in a (perhaps) more plausible
scenario where the product of star formation efficiency and escape
fraction for Pop-III stars is comparable to that for Pop-II stars,
$\tau_{\rm es}=0.09-0.12$ would be observed [consistent with our
previous calculations (Cen~2003a,b)]. In this case the histories would
be characterized by an extended ionization plateau from $z=7-12$
where the mean neutral fraction stays in a narrow range of $0.1-0.3$.
These results are rather insensitive to the redshift where the IGM
becomes enriched with metals.

\acknowledgements 
We thank Jerry Ostriker for helpful discussions.
This research is supported in part by the Australian Research Council, 
and by grants AST-0407176 and NAG5-13381.

\end{document}